\documentclass[10pt]{article}
\usepackage{stmaryrd}
\usepackage{amsmath}
\usepackage{amssymb}
\usepackage{url}
\usepackage[utf8]{inputenc}

\usepackage{graphicx}

\usepackage{cite}

\usepackage{color} 

\usepackage[labelfont=bf,labelsep=period,justification=raggedright]{caption}

\makeatletter
\renewcommand{\@biblabel}[1]{\quad#1.}
\makeatother

\date{}

\begin{document}

\begin{flushleft}
{\Large
\textbf{Detecting the community structure and activity patterns of temporal networks: a non-negative tensor factorization approach}
}
\\
Laetitia Gauvin$^{1\ast}$, 
Andr\'e Panisson$^{1}$,
Ciro Cattuto$^{1}$
\\
\bf{1} Data Science Laboratory, ISI Foundation, Torino, Italy
\\
$\ast$ email: laetitia.gauvin@isi.it
\end{flushleft}
\section*{Abstract}

The increasing availability of temporal network data is calling for more research
on extracting and characterizing mesoscopic structures in temporal networks
and on relating such structure to specific functions or properties of the system.
An outstanding challenge is the extension of the results achieved for static networks to time-varying networks,
where the topological structure of the system and the temporal activity patterns of its components
are intertwined.

Here we investigate the use of a latent factor decomposition technique, non-negative tensor factorization,
to extract the community-activity structure of temporal networks. The method is intrinsically temporal
and allows to simultaneously identify communities and to track their activity over time. 
We represent the time-varying adjacency matrix of a temporal network as a three-way tensor
and approximate this tensor as a sum of terms that can be interpreted as communities of nodes
with an associated activity time series.
We summarize known computational techniques for tensor decomposition
and discuss some quality metrics that can be used to tune the complexity of the factorized representation.

We subsequently apply tensor factorization to a temporal network
for which a ground truth is available for both the community structure and the temporal activity patterns.
The data we use describe the social interactions of students in a school, the associations between students and school classes, and the spatio-temporal trajectories of students over time.
We show that non-negative tensor factorization is capable of recovering the class structure
with high accuracy. In particular, the extracted tensor components can be validated either as known school classes,
or in terms of correlated activity patterns, i.e., of spatial and temporal coincidences that are determined by the known school activity schedule.



\section*{Introduction}


Many natural and artificial systems can be fruitfully represented as networks involving elementary structural entities and specific relations between them. Among the insights that the network representation can provide, a central aspect is the relationship between network structure and system's function. To this end, a great deal of work has been devoted to detecting and identifying clusters or communities in static networks, assessing their statistical relevance, and linking community structure to network function~\cite{Fortunato}.
Most real-world network systems, however, are in constant evolution, and the increasing availability of time-resolved network data sources, e.g., from socio-technical systems and on-line social networks, has brought to the forefront the need to study and understand time-varying networks~\cite{Holme201297}.
Although it is always possible to create static network representations by aggregating over the temporal evolution of the system, such temporally-aggregated representations may overlook essential features of the system or may confound structures that can be teased apart only by retaining the time-varying nature of the data. For example, a node or group of nodes may belong to different communities at different points in time: aggregating the network over time will artificially merge those communities and create a cluster that does not represent the reality of the system at any point in time. Similarly, groups of nodes may exist that share similar activity patterns over time (due to, e.g., an externally imposed activity schedule): an aggregated view on the network will only retain the topology of the interactions and lose the activity patterns and temporal correlations.
Overall, detecting structures that involve topological features and correlated activity patterns over time is an outstanding challenge that bears relevance to many fields of research and needs a principled approach as well as efficient computational methods.

Recent work addressed the community detection problem on time-varying networks by finding communities in snapshots of the networks at different times and then analyzing the changes of the community structures and linking the structures found at different times.
Simple approaches to mine the time-varying community structure of a system~\cite{chen2013detecting,Hopcroft,Greene2010} are based on a continuity assumption for the (static) 
community structure detected at successive time intervals. These approaches may prove useful in specific cases, but fail in the case of discontinuous activity patterns, abrupt structure formation or dissolution, 
and in general they cannot deal with temporal correlations over extended periods of time.
Instead of treating separately the community structure and the temporal evolution of the network, a few studies~\cite{Bassett,Mucha,Ronhovde} pioneered global approaches to community detection in temporal networks, even though the current lack of benchmarks makes the evaluation of these methods difficult.

Here we propose a method to detect the community-activity structure of temporal networks, and we validate this method using an empirical temporal network for which a ground truth is available for both the community structure and the temporal activity patterns. 
The method we study is intrinsically temporal and allows us to simultaneously identify communities and to track their activity over time.
It is based on the fact that a temporal network is naturally represented as a time-ordered sequence of adjacency matrices, each one describing the state of the network at a given point in time. The adjacency matrices can be combined in a three-way tensor, for which a number of mathematical techniques for multi-layer networks can be applied~\cite{moreno2013}, together with established methods from data mining and machine learning~\cite{Cichocki:1253894,Morup11}.
The approach described here is based on tensor factorization techniques that were developed to extract latent signals in diverse domains like signal processing, psychometrics, brain science, linguistics and chemometrics~\cite{kolda,Shashua:2005:NTF:1102351.1102451,Cichocki:1253894,VandeCruys:2009:NTF,Beyondstreams,Wang:2011}. We base our work on the so-called canonical decomposition~\cite{Candecom}, also known as parallel factorization~\cite{harshman1970fpp}, which can be regarded as a generalization of singular value decomposition (SVD) to tensors.
In particular, we focus on non-negative tensor factorization~\cite{Shashua:2005:NTF:1102351.1102451,Cichocki:1253894}, since -- as already observed for non-negative matrix factorization \cite{lee1999learning} -- it is a powerful tool for learning 
parts-based representation of a dataset, resulting in more interpretable models \cite{Dunlavy, ICML2011Nickel_438}.
Non-negative factorization techniques have been already proposed for community detection in static networks~\cite{Wang:2011-NMF,Yang:2013} because of their ability to capture densely overlapping communities.

A central challenge in designing techniques for structure detection and extraction is the ability to validate the obtained results by comparing them with an externally available ground truth. Here we leverage a very particular dataset on time-resolved social interactions in a school, for which the full class structure of the school and the activity schedule of the classes are independently available.
This dataset represents an interesting case study, as there are structures at different scales, both topologically and temporally, that arise from the spatial, social, and temporal dimensions of the school activity. We apply our factorization technique to the tensor describing the temporal network of social contacts and  extract the time-varying community structure of the empirical data.
We show that our method fully recovers the known class structure of the school, the activity patterns of classes over time, and it also detects communities spanning mixed classes that correspond to known social activities in the public spaces of the school.

\section*{Materials and Methods}

\subsection*{Empirical temporal network data}

In order to enable the validation of our results, we leverage a high-resolution dataset that describes the close-range social interactions of children in a primary school~\cite{plosOne}. The data were collected by the SocioPatterns collaboration\footnote{http://www.sociopatterns.org} using wearable proximity sensors that sense the face-to-face proximity relations of individuals wearing them.
This dataset presents expected structures (classes) as well as potential less conspicuous structures. This thus appears as an ideal dataset to assess the efficiency of the NTF community detection algorithm. This contrasts with the fact that, until now, no clear benchmark has been established to estimate the quality of community detection algorithms on time-varying networks.

\textbf{Temporal social network.}
The population of the school consisted of $241$ children aged $6$ to $12$ and organized in $10$ classes, together with $10$
teachers. Each participant was equipped with a badge containing a proximity sensor with a unique identifier. The sensor continuously monitored the close-range
(less than $1-1.5$ meters) face-to-face contacts of individuals and relayed the proximity relations to a receiving system that timestamps and logs the
data~\cite{Cattuto2010}. The data were collected over two consecutive days in October 2009 from 8:30am to 5:15pm and only interactions taking place on the
premises of the school were recorded. The system has a temporal resolution of $20$ seconds, so that proximity relations are detected over consecutive $20$-second
time intervals. The empirical data are therefore naturally represented as a temporal social network. The data from $10$ sensors were discarded because of data
quality reasons, thus in the following we will work with a temporal network with $N=241$ nodes.

\textbf{Class structure of the population.}
No personal information is associated with the unique identifiers of the wearable sensors. However, each identifier is associated to the class the participant belongs 
to, so that we have a ground truth for the communities that define the class structure of the monitored population. Table S$1$
in the Supplementary Information reports summary information on the school classes.

\textbf{Spatial information.}
The radio packets transmitted by the wearable proximity sensors were picked up by $15$ receivers (readers) located throughout the
school grounds. For a given sensor $s$, the number of packets received by a reader is a decreasing function of the distance between that sensor and the reader.
The number of packets per unit time received by the readers can thus be regarded as a spatial ``fingerprint'' that provides information on the location of
sensors with a room-level accuracy.
We define a spatial feature vector for all sensors by counting, for each sensor $s$, the number of packets $f_{i}^{(s)}$ received by receiver $i$ over a fixed time interval. Here we choose to aggregate the location information over consecutive $1$-minute interval.
We thus represent the location fingerprint of tag $s$ at time $t$ using the vector $(f_{1}^{(s)}(t), ..., f_{n}^{(s)}(t))$, with $n=15$ receivers.
The availability of spatial information over time allows us to define trajectories for individual tags as well as for groups of tags with the same class label.

\subsection*{Tensor representation of the empirical data}
The temporal network dataset we use comprises two days of recorded social interactions with a temporal resolution of $20$ seconds.
The  schedule of classes and social activities that we use as a ground truth for the activity timelines, however,
is defined on a coarser temporal scale. Hence for the present study we aggregate the raw sensor data
over longer time intervals, comparable to this temporal scale. 
Different levels of aggregation can be chosen, according to the temporal scale of the activity timelines to be explored.

In what follows, we divide the dataset timeline into $S = 150$ consecutive intervals of
approximately $13$ minutes, and we aggregate the temporal network for each interval.
We also considered intervals of $5, 15, 30, 60$ minutes,
which are comparable to the typical temporal scale of activities at school, to study the robustness of the results with regards to the choice of aggregation level
(details of the comparison are found in the Supplementary Information).
The division of the total duration of the experiment in $150$ intervals and the subsequent aggregation yields 150 network snapshots,
built so that one link is drawn between two nodes if those nodes had at least
one contact during the corresponding interval.
The state of a network during one interval is represented by an adjacency matrix $\textbf{M} \in \mathbb{R}^{N \times N}$,
where the binary-valued entry $M_{ij}$ indicates the presence of the $i$-$j$ link.
The temporal network can thus be represented as $150$ successive adjacency matrices
combined into a 3-way tensor, $\mathcal T \in \mathbb{R}^{N \times N \times S}$.


\subsection*{Uncovering latent structures by tensor factorization}
\begin{figure}[!htbp]
\begin{center}
\includegraphics[width=12cm]{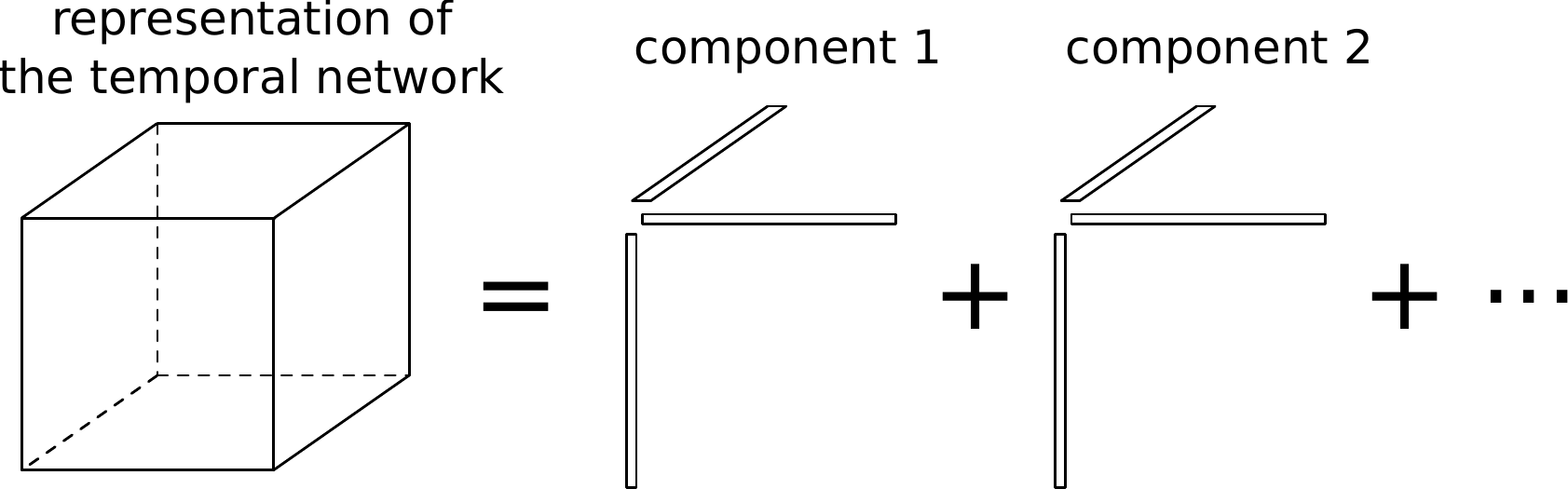}
 \end{center}
 \caption{{\bf Pictorial representation of the Kruskal decomposition.} The cube on the left is the original 3-way tensor, which is represented as the sum of rank-1 tensors (on the right), each generated as the outer product of three 1-dimensional vectors (thin rectangles). Each of the rank-1 terms on the right corresponds to one component.}
\label{drawing_factorization}
\end{figure}

The tensor $\mathcal T \in \mathbb{R}^{N \times N \times S}$, where $N$ is the number of nodes of the network
and $S$ the number of network snapshots, encodes both the topological and temporal information on the network under study.
Uncovering structures that may correspond to communities or correlated activity patterns requires the identification
and extraction of lower-dimensional factors. To this end, we use tensor factorization techniques,
i.e., we choose to represent the tensor as a suitable product of lower-dimensional factors.
This can be achieved by means of the so-called canonical decomposition (canonical polyadic decomposition, CP).
CP in $3$ dimensions aims at writing a tensor $\mathcal T \in \mathbb{R}^{N \times N \times S}$ in a factorized fashion:
 \begin{equation}
 t_{ijk}=\sum_{r=1}^{R_\mathcal{T}} \, a_{ir} b_{jr} c_{kr} \, ,
 \label{element-approx}
\end{equation}
where the smallest value of $R_\mathcal{T}$ for which such a relation can hold is the rank of the tensor $\mathcal{T}$.
In other words, the tensor $\mathcal{T}$  can always be expressed as a sum of rank-$1$ tensors in the form
\begin{equation}
\mathcal{T} = \sum_{r=1}^{R_\mathcal{T}}  \, \mathbf{a_r} \circ \mathbf{b_r} \circ \mathbf{c_r}  \, ,
\label{approx}
\end{equation}
i.e., as the sum of outer products of three vectors.
The set of vectors
$a_{\{1,2,\dots,R_\mathcal{T}\}}$ (resp. $b_{\{1, 2, \dots, R_\mathcal{T}\}}$, $c_{\{1, 2, \dots, R_\mathcal{T}}\}$)
can be re-written as a matrix $\mathbf{\hat{A}} \in \mathbb{R}^{N \times R_\mathcal{T}}$ 
(resp. $\mathbf{\hat{B}} \in \mathbb{R}^{N \times R_\mathcal{T}}$ and $\mathbf{\hat{C}} \in \mathbb{R}^{S \times R_\mathcal{T}}$),
where each of the $R_\mathcal{T}$ vectors is a column of the matrix.
The decomposition of Eq.~\ref{approx} can therefore be represented in terms of the three matrices
$\mathbf{\hat{A}}, \mathbf{\hat{B}}, \mathbf{\hat{C}}$
as $\llbracket \mathbf{\hat{A}}, \mathbf{\hat{B}}, \mathbf{\hat{C}} \rrbracket$.
A visual representation of this factorization, also known as the Kruskal decomposition,
is shown in Fig.~\ref{drawing_factorization}.

\subsection*{Factorization methodology}
In the present case, each rank-$1$ tensor, that we henceforth call \textit{component}, corresponds to a set of nodes whose activities are correlated.
The aim here is not to find an exact factorization, but rather to approximate the tensor with a number of components smaller than the rank of the original tensor.
Such an approximation of the tensor is equivalent to minimizing the difference between
$\mathcal{T}$ and $\llbracket \mathbf{A}, \mathbf{B}, \mathbf{C} \rrbracket$ (PARAFAC decomposition),
\begin{equation}
\min\limits_{\mathbf{A},\mathbf{B},\mathbf{C}} {\Vert \mathcal{T} -\llbracket \mathbf{A},\mathbf{B},\mathbf{C} \rrbracket \Vert}^2_F \, ,
\label{3d-problem}
\end{equation}
where $\mathbf{A}, \mathbf{B}, \mathbf{C}$ respectively have dimensions $N \times R$, $N \times R$ and $S \times R$,
$R<R_\mathcal{T}$, and $\Vert A \Vert^2_F = \sqrt{\sum_{ijk}|a_{ijk}|^2}$ is the Frobenius norm.
In the following $\mathbf{A}, \mathbf{B}, \mathbf{C}$ will always indicate the approximate decomposition, to avoid confusion with the exact decomposition $\mathbf{\hat{A}}, \mathbf{\hat{B}}, \mathbf{\hat{C}}$ mentioned above.

Solving this problem amounts to finding the $R$ rank-1 tensors that best approximate the tensor $\mathcal{T}$.
The number $R$ of components is chosen on the basis of the desired level of detail:
a low number of components only yields the strongest structures, potentially overlooking important features,
whereas using a high number of components faces the risk of overfitting noise.
Choosing $R$ amounts to an optimization problem in which we seek the number of components that best explain
the structure of the tensor without describing the possible noise of the data.
In this respect, the tensor factorization method is similar to community detection techniques where the number
of communities is fixed a priori: the number of components we choose to approximate the tensor
is the number of communities or activity patterns we extract (see also Fig.~\ref{drawing_factorization}).

We transform the $3$-dimensional problem of Eq.~\ref{3d-problem} into $2$-dimensional sub-problems by unfolding the tensor $\mathcal{T}$ through a process called
matricization: The mode-$i$ matricization consists in linearizing all the indices of the tensor except $i$. In our case this yields three modes:
$\mathbf{X}_{(1)}, \mathbf{X}_{(2)}, \mathbf{X}_{(3)}$. The three resulting matrices have respectively a size of $N \times NS$, $N \times NS$ and $S\times N^2$.
Each element of the matrix $\mathbf{X}_{(i=1,2,3)}$ corresponds to one element of the tensor $\mathcal{T}$, i.e., each mode contains all the values of the
original tensor.
Thanks to matricization, the factorization problem of Eq.~\ref{approx} can be reframed in terms of individual factorizations of the three modes. In other words, minimizing the difference between
$\mathcal{T}$ and $\llbracket \mathbf{A},\mathbf{B},\mathbf{C} \rrbracket$
is equivalent to minimizing  the difference between each of the modes and their respective approximation in terms of $\mathbf{A}, \mathbf{B}, \mathbf{C}$:
\begin{eqnarray}
\mathbf{X}_{(1)} = \mathbf{\hat{A}}(\mathbf{\hat{C}}\odot \mathbf{\hat{B}})^T \nonumber \\
\mathbf{X}_{(2)} = \mathbf{\hat{B}}(\mathbf{\hat{C}}\odot \mathbf{\hat{A}})^T \\
\mathbf{X}_{(3)} = \mathbf{\hat{C}}(\mathbf{\hat{B}}\odot \mathbf{\hat{A}})^T \nonumber
\end{eqnarray}
where $\odot$ denotes the Khatri-Rao product, which is a column-wise Kronecker product,
i.e., $\mathbf{C}  \odot \mathbf{B}  = [c_1 \otimes b_1, c_2 \otimes b_2,  \dots, c_r \otimes b_r]$.
If $\mathbf{C} \in \mathbb{R}^{S \times R}$ and $\mathbf{B} \in \mathbb{R}^{N \times R}$,
then the Khatri-Rao product $\mathbf{C} \odot \mathbf{B} \in \mathbb{R}^{SN \times R}$.
Overall, the factorization problem of Eq.~\ref{3d-problem} (PARAFAC) is  converted into the three following sub-problems:
\begin{eqnarray}
\min\limits_{\mathbf{A}\geq 0} {\Vert \mathbf{X}_{(1)} -\mathbf{A}(\mathbf{C}\odot \mathbf{B})^T \Vert}^2_F \nonumber \\
\min\limits_{\mathbf{B} \geq 0} {\Vert \mathbf{X}_{(2)} -\mathbf{B}(\mathbf{C}\odot \mathbf{A})^T \Vert}^2_F \\
\min\limits_{\mathbf{C} \geq 0} {\Vert \mathbf{X}_{(3)} -\mathbf{C}(\mathbf{B}\odot \mathbf{A})^T \Vert}^2_F \nonumber
\label{2d-problem}
\end{eqnarray}
Here we focus on \textit{non-negative} factorization, i.e., we impose a condition of non-negativity
on all the elements of the three modes.
This is customarily used to achieve a purely additive representation of the tensor in terms of components,
which greatly simplifies the interpretation of the resulting decomposition~\cite{lee1999learning}.

In the case of temporal networks, $\mathbf{A},\mathbf{B},\mathbf{C}$, also called factors, give access to different interpretations:
$\mathbf{A}$ and $\mathbf{B}$ provide the community structure of the network
and $\mathbf{C}$ gives the temporal activity of each community.
For an undirected network the adjacency matrix represented on each tensor slice is symmetric,
and $\mathbf{A}=\mathbf{B}$. In this case, the result of tensor factorization is illustrated in Fig.~\ref{fact_result}.
\begin{figure}[!htbp] 
\begin{center}
     \includegraphics[width=7cm]{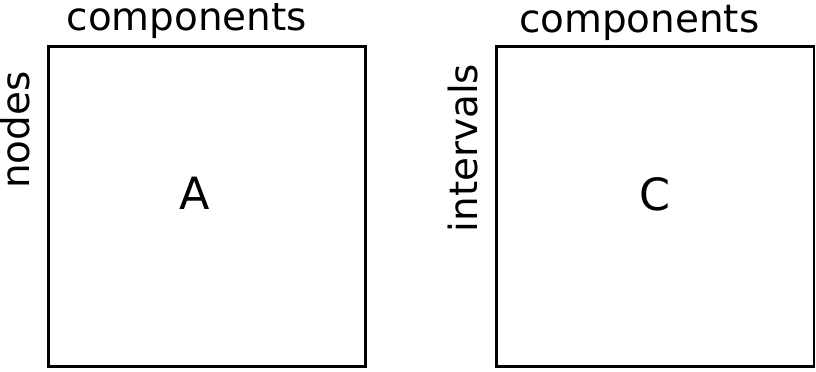}
\end{center}
 \caption{{\bf Schematic representation of the factorization result for an undirected temporal network.}
 The factors $\mathbf{A}$ and $\mathbf{C}$ are matrices with $R$ columns, each one corresponding to one extracted component.
 The rows of $\mathbf{A}$ correspond to network nodes, and the rows of $\mathbf{C}$  to discrete time intervals.
 The entries of $\mathbf{A}$ give the membership weight of nodes to the different components.
 The entries of $\mathbf{C}$ give the activity level of components at different intervals.
 }
\label{fact_result}
\end{figure}

Several algorithms have been developed to carry out the PARAFAC decomposition describe above.
The two most common techniques are the projected gradient method~\cite{Lee00algorithmsfor} and the alternating least squares (ALS) method. In the ALS method~\cite{Phan2012-1} the $3$ problems of Eq.~\ref{2d-problem} are solved by alternating
a minimization procedure in which two of the matrices are kept fixed while the third is varied for minimization.
The tensor factorization technique we use here is based on the non-negative alternate least squares method (ANLS \cite{Paatero}) combined with a block-coordinate-descent technique~\cite{Bertsekas1999,Park-NTF},
that achieves faster convergence. Our implementation uses the Tensor Toolbox \cite{Tensor-Toolbox}.

\subsection*{Assessing the quality of factorization: how to control a multi-scale method}
Increasing the number $R$ of components allows us to represent more and more structure of the temporal network. However, as the number of components increases, we go from underfitting to overfitting these structures, i.e., we face the usual trade-off between approximating complex structures and overfitting them, potentially capturing noise.
This is a characteristic of any intrinsically multi-scale method, and it is important to control it by designing and using quality metrics for the obtained decompositions that can guide their use in the context of a specific research question or application.
Notice that we do not aim at setting an ``optimal'' number of components, but rather at assessing the quality of a decomposition obtained for a given choice of $R$. Here we make use of a standard metric called ``core consistency''~\cite{Bro}, which we briefly describe in the following.

We notice that the tensor decomposition of Eq.~\ref{element-approx} can be written as 
\begin{equation}
 t_{ijk} = \sum_{m=1}^{R_\mathcal{T}}  \sum_{n=1}^{R_\mathcal{T}} \sum_{p=1}^{R_\mathcal{T}} \, g^{1}_{mnp} \, a_{im} b_{jn} c_{kp} \, ,
 \label{tucker-kruskal-approx}
\end{equation}
where $g^{1}_{ijk}$ is the unit superdiagonal tensor. This form is a special case of a more general tensor decomposition known as Tucker decomposition:
\begin{equation}
 t_{ijk} = \sum_{m=1}^{M}  \sum_{n=1}^{N} \sum_{p=1}^{P} \, g_{mnp} \, a_{im} b_{jn} c_{kp} \,\, ,
 \label{tucker-approx}
\end{equation}
where the tensor $g_{mnp}$, known as core tensor,  encodes the interactions between the three factors.
It can be shown that for a perfectly-fitted PARAFAC model, yielding factors $\mathbf{A}$, $\mathbf{B}$ and $\mathbf{C}$,
the core tensor of the Tucker decomposition obtained by fixing the factors and minimizing over $g_{mnp}$
is the unit super-diagonal tensor $g^1$ (if the factors have full column rank, see Ref.~\cite{Bro}).
This points to a possible way to assess the appropriateness of a PARAFAC tensor decomposition with $R$ components:
we first fix $R$ and compute the PARAFAC decomposition with $R$ components, obtaining the factors
$\mathbf{A}$, $\mathbf{B}$, and $\mathbf{C}$. Then we compute the Tucker decomposition of Eq.~\ref{tucker-approx}
with $M=R$, $N=R$, $P=R$ and the factors $a_{im}$, $b_{jn}$, and $c_{kp}$ fixed to the result of the former PARAFAC decomposition, obtaining the core tensor $g_{mnp}$. Finally, we compare $g_{mnp}$ with the unit super-diagonal tensor $g^1$,
and quantify their similarity by a metric known as ``core consistency'':
\begin{equation}
CC(R) = 1\, - \, \frac{1}{R}  \sum_{m=1}^{R} \sum_{n=1}^{R}  \sum_{p=1}^{R}  \, (g_{mnp} - g^{1}_{mnp})^2 \,\, .
\label{eq:coreconsistency}
\end{equation}
Typically, on plotting the value of $CC(R)$ for an increasing number of components $R$, a crossover can be observed between high core consistency values for low
$R$ and lower core consistency values for high $R$, when the PARAFAC model with $R$ components stops being a proper description of the original tensor because it
overfits or the components become redundant. Values of $CC(R)$ greater $0.5$ are generally considered acceptable~\cite{Bro}, and the value of $R$ for which
$CC(R)$ crosses over is usually used as a guide for setting the optimal range for the number of components.

Given a choice of $R$, a complementary way to estimate the quality of a specific PARAFAC decomposition with $R$ components is to quantify how much of the
original signal is recovered by the extracted components. To this end, we start by quantifying the weight that a given component $r$ has in each of the factors
$\mathbf{A}$, $\mathbf{B}$, $\mathbf{C}$: we compute the L2-norm of the $r$th column of each factor, yielding the norms $\lambda^A_r$, $\lambda^B_r$ and
$\lambda^C_r$ and we define the relevance of component $r$ as the product of these norms, $\lambda_r = \lambda^A_r \cdot \lambda^B_r \cdot \lambda^C_r$. This
allows us to rank the components by the contribution they give to the decomposition, and to score a whole decomposition by the product over $r$ of its $\lambda_r$.

Finally, it is important to anticipate that, as reported in the Supplementary Information, the community structures
and activity patterns we obtain on our dataset are very robust with respect to the number of components:
on increasing (or decreasing) $R$, new structures are uncovered (or lost), but most components
stay stable both in terms of topology and in terms of activity patterns.

\subsection*{Interpreting the factors: community structure and activity patterns}
The factor matrices $\mathbf{A}, \mathbf{B}, \mathbf{C}$ all have $R$ columns, each of them corresponding to one component.
Since we used non-negative tensor factorization, all entries of these matrices are non-negative.
In the special case of an undirected network, $\mathbf{A} = \mathbf{B}$. The elements $a_{ir}$ of matrix $\mathbf{A}$
associate each component $r$ to the nodes $i$ it spans, i.e., they describe community structure of the original network,
with the matrix entries providing weights for the membership of nodes to such communities.
The elements $c_{kr}$ of matrix $\mathbf{C}$, on the other hand, associate each component $r$ to the time intervals $k$ it spans,
and the matrix values for a given component indicate the activity level of that component as a function of time (index $k$),
i.e., its temporal activity pattern.

We remark that individual nodes can be members of different components, with different weights. That is, non-negative factorization of the temporal network
tensor can naturally capture overlapping communities. This mirrors the results of the study by Yang and Leskovec~\cite{Yang:2013}, where non-negative matrix
factorization was shown capable of detecting densely overlapping as well as non-overlapping communities in static networks. Similarly, non-negative tensor
factorization allows to extract non-overlapping temporal communities, densely overlapping communities, and multi-scale community structure.

As noted above, the factor $\mathbf{C}$ yields the temporal activity of each component (community), irrespective of the node composition of the component. We
define the activity level of each community $r$ at a given interval $k$ (time index) by using both the information on the temporal activity of the component
(from $\mathbf{C}$) and the memberships strength of each node (from $\mathbf{A}$) in that component. The strength $s_{kr}$ of a component $r$ at interval $k$ is
therefore defined as:
\begin{equation}
s_{kr}=c_{kr} \sum_{i=1}^{N} a_{ir} \, ,
\end{equation}
where $r=1,\dots,R$ and $k=1,\dots,S$.
Notice that the activity of a component over time can be very uneven, i.e., it is possible to capture structures (components) that have temporally-disjoint
activity regions. This is a consequence of the fact that non-negative tensor factorization captures purely structural aspects of the original network tensor and
does not rely or impose in any way constraints of temporal continuity on the detected structures.

\begin{figure}[!htbp] 
\begin{center}
     \includegraphics[width=9cm]{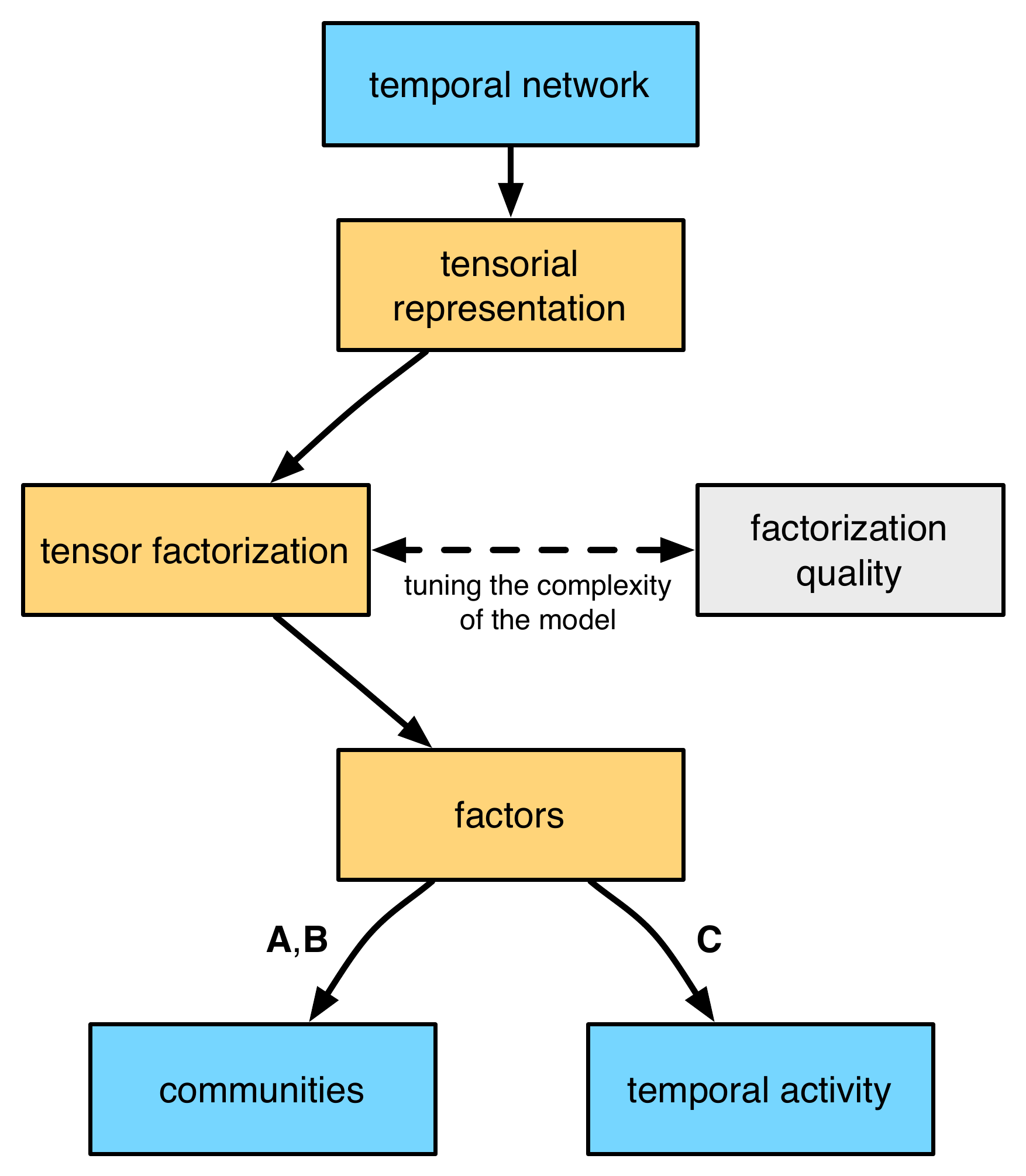}
\end{center}
 \caption{{\bf Community-activity structure detection via non-negative tensor factorization.}
 The original temporal network is represented as a three-way tensor, which is then decomposed by using non-negative tensor factorization. The complexity of the model (number of components $R$) is tuned by using quality indicators
 that provide information on the stability, coverage or redundancy of the decomposition.}
\label{Fig:process}
\end{figure}

To summarize the structure detection technique we described above, Fig.~\ref{Fig:process} schematically illustrates
the roles played by the tensorial representation, non-negative tensor factorization,
core consistency analysis, and the interpretation of matrix factors.

\section*{Results}
In the following we report the results obtained by applying the structure detection methodology of Fig.~\ref{Fig:process}
to the empirical temporal network of social interactions described in the Materials and Methods section.

\subsection*{Factorizing a high-resolution temporal social network}
The empirical network data describes the close-range interactions of $231$ students and $10$ teachers, divided in $10$ classes.
The total duration of the experiment was segmented in $150$ intervals of $13$-minutes for the purpose of the analysis.
We build the tensorial representation of the temporal network and compute the non-negative factorization as described in the Methods section, for this case and other aggregation levels ($5,15,30,60$ minutes) as well.
Here we describe the case of $150$ intervals but the other aggregation levels give comparable results as detailed in the Supplementary Information.
We approximate the empirical tensor as the sum of $R$ components, 
according to Eq.~\ref{element-approx}, with $R$ ranging from $2$ to $20$. Since the optimization problem does not have a unique global minimum, for each value of $R$ we ran the optimization method $20$ times, 
each time starting with different initial condition for the factors $\mathbf{A}$ and $\mathbf{C}$, and computed the core consistency of Eq.~\ref{eq:coreconsistency} for each run. For each choice of $R$ we rank the runs by their core
consistency and we select the top $50\%$ runs. Out of these, we select the top $5$ decompositions with the highest sum of component weights (the $\lambda_r$ described above). The corresponding core consistency values are plotted as a
function of $R$ in Fig.~\ref{Fig:Com} (left), where an abrupt change in the slope is visible for a critical value $R \simeq \hat{R} = 13$.

This change of slope indicates that for $R > \hat{R}$ most of the intrinsic structures of the dataset have been captured, the obtained decompositions may start overfitting and hence risk being less stable with respect to noise and initial conditions.
Conversely, all decompositions obtained for $R \lessapprox \hat{R}$ yield core consistency values in excess of $0.9$, which is regarded as an indicator of robust structures captured by the factorization method~\cite{Bro}. In this regime, different 
number of components yield different levels of structural detail of the tensor.
We remark that the observed behavior is independent of the above choices on the number of factorization runs, and of the specific thresholds used for selecting the best ones.

In the following, in order to discuss the structures detected by the method and to validate them in terms of our ground truths, we focus on a specific decomposition with $R=\hat{R}=13$. 
This choice, guided by the core consistency curve, corresponds to selecting the most
complex models that yields a robust decomposition.

\subsection*{Community structure and activity patterns}

The temporal social network we study is undirected, hence the factors $\mathbf{A}$ and $\mathbf{B}$ are identical and provide the membership scores that associate network nodes to the different components ($R=13$) extracted 
by non-negative factorization,
as discussed in the Materials and Methods section.
In the specific case of our dataset the weights $a_{ir}$ exhibit a strong peak at $a_{ir} = 0$ and the other weights are distributed more broadly around a non-zero value.
A typical distribution of membership weights is displayed in Fig.~\ref{Fig:Bimodal}, 
and the distributions for all the $13$ components we consider are reported in the Supplementary Information. 
For each component $r$, this allows us to naturally divide the weights in two classes, i.e., to classify the network nodes
as member or non-members of a given component (e.g., by using an unsupervised clustering technique such as k-means
with two clusters) in a robust fashion. The memberships to the components can be summarized in a node-component matrix $\tilde{\mathbf{A}}$. The terms $\tilde{a}_{ni}$ of the node-component matrix  
are $1$ (resp. $0$) if the node $n$ is member (resp. not member) of the component $i$.
The only nodes which can alternatively be classified as member of the community or outside of it depending on the clustering technique
are those with small activity compared to the others, but such fluctuations concern few individuals with regard to the total size of the community.
We remark that this is not necessary in order to make use of the (weighted) community structure information contained in the factor $\mathbf{A} = \mathbf{B}$, and here we proceed to classify each node as 
member (or not member) of a given component (community) simply because our dataset allows this, and the resulting binary classification affords a simpler representation, analysis, and validation of the structures we find. 
Less structured temporal networks, in general, should not be expected to yield membership weights that can be cleanly separated in two classes.
\begin{figure}[!htbp]
 \begin{center}
     \includegraphics[width=8cm]{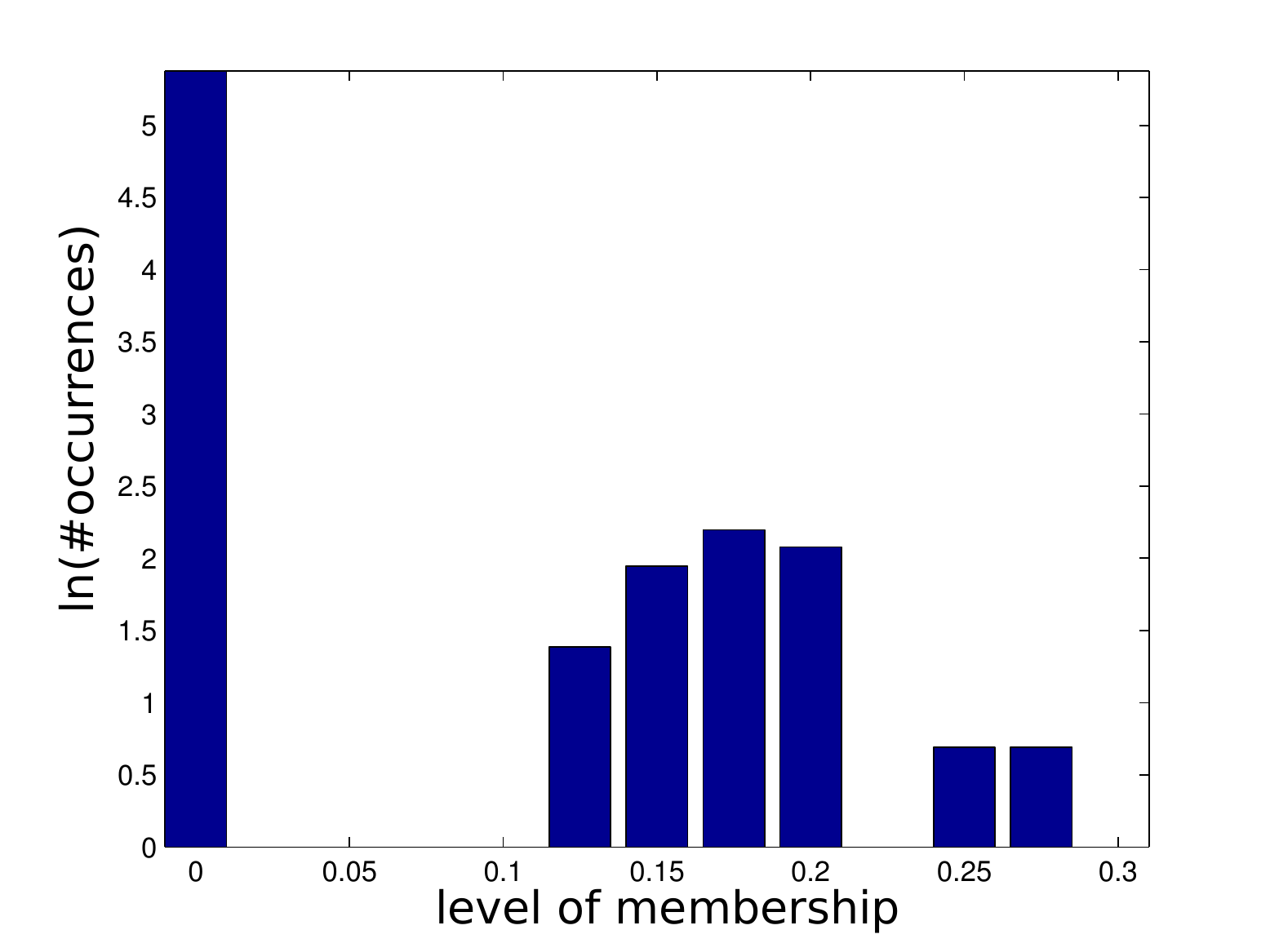}
\end{center}
\caption{{\bf Distribution of the membership weights.}
Sample histogram of the membership weights for one component of the decomposition
(one column of factor $\mathbf{A}$ for $R=13$).
}
\label{Fig:Bimodal}
\end{figure}

\begin{figure}[!htbp]
\begin{center}
   \begin{minipage}[c]{.44\linewidth}
      \includegraphics[width=7cm]{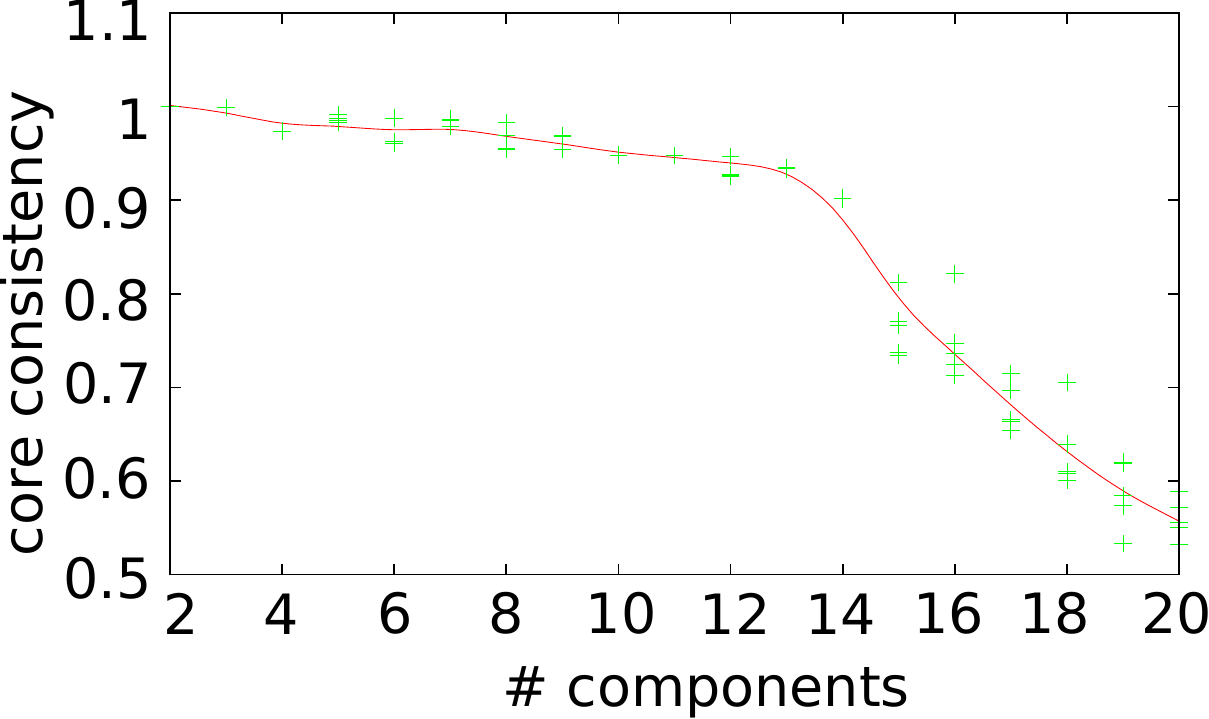}
   \end{minipage} \hfill
   \begin{minipage}[c]{.44\linewidth}
      \includegraphics[width=7cm]{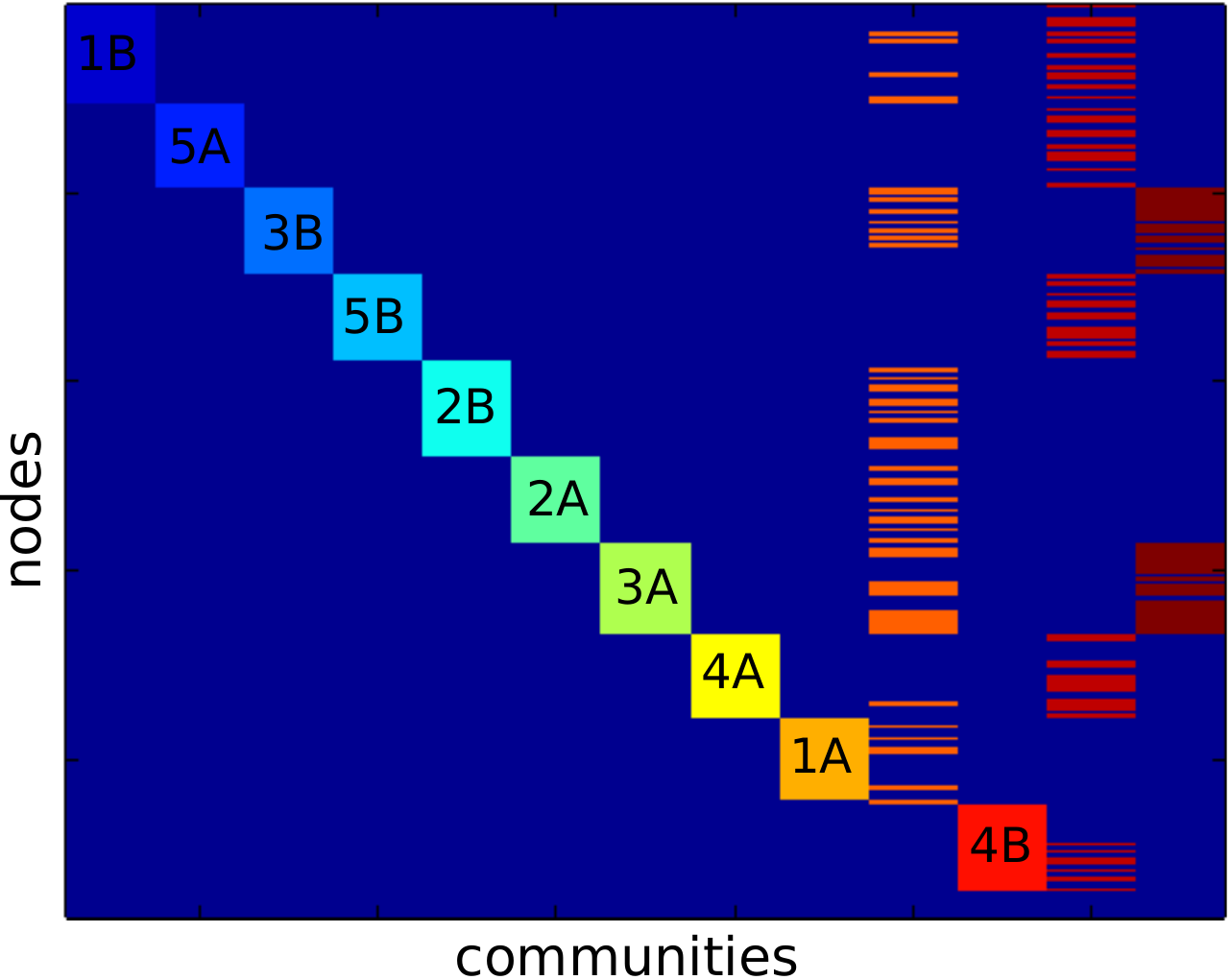}
   \end{minipage}
\caption{{\bf NTF decomposition of an empirical temporal network.}
Left panel: core consistency curve. For each value of the number of components $R$ used for factorization, the core consistency values for the 5 best decompositions are reported (crosses). The solid line is a guide for the eye. A crossover between two regimes is visible for $R \simeq 13$.
Right panel: component-node matrix for $R=13$ components. Rows correspond to network nodes and columns to components. The matrix is obtained from the factor $\mathbf{A}$ by classifying each node as belonging (lighter rectangles) or not belonging (dark blue rectangles) to a given component. The order of the nodes has been rearranged to expose the block structure of the matrix. Colors identify components, and the community structures that can be
matched to school classes are annotated with the corresponding class name. }
\label{Fig:Com} 
\end{center}
\end{figure}

The right-hand panel of Figure~\ref{Fig:Com} summarizes the community structure detected by non-negative tensor factorization, displaying the binary association between nodes (along the vertical axis) and components (along the horizontal axis). Components are color-coded, and the order of nodes along the vertical axis has been adjusted to expose the strong block structure of the matrix plot.
The $10$ blocks correspond to mutually disjoint communities, whereas the remaining $3$ components (the two rightmost components and the fourth component from the right) are larger and have a significant overlap with one another and with said $10$ communities, mixing from $2$ to $7$ of them. The sizes of the detected communities are reported in Table S2 
of the Supplementary Information. 
We anticipate that the $10$ mutually disjoint communities correspond to the $10$ classes of the school,
as we will discuss in the validation section below.
Most of the nodes are found to belong to at least one community ($234$ out of $241$). The remaining nodes, on direct inspection of the dataset, have a negligible activity and they are not part of any community.

\begin{figure}[!htbp] 
\begin{center}
     \includegraphics[width=0.9\textwidth]{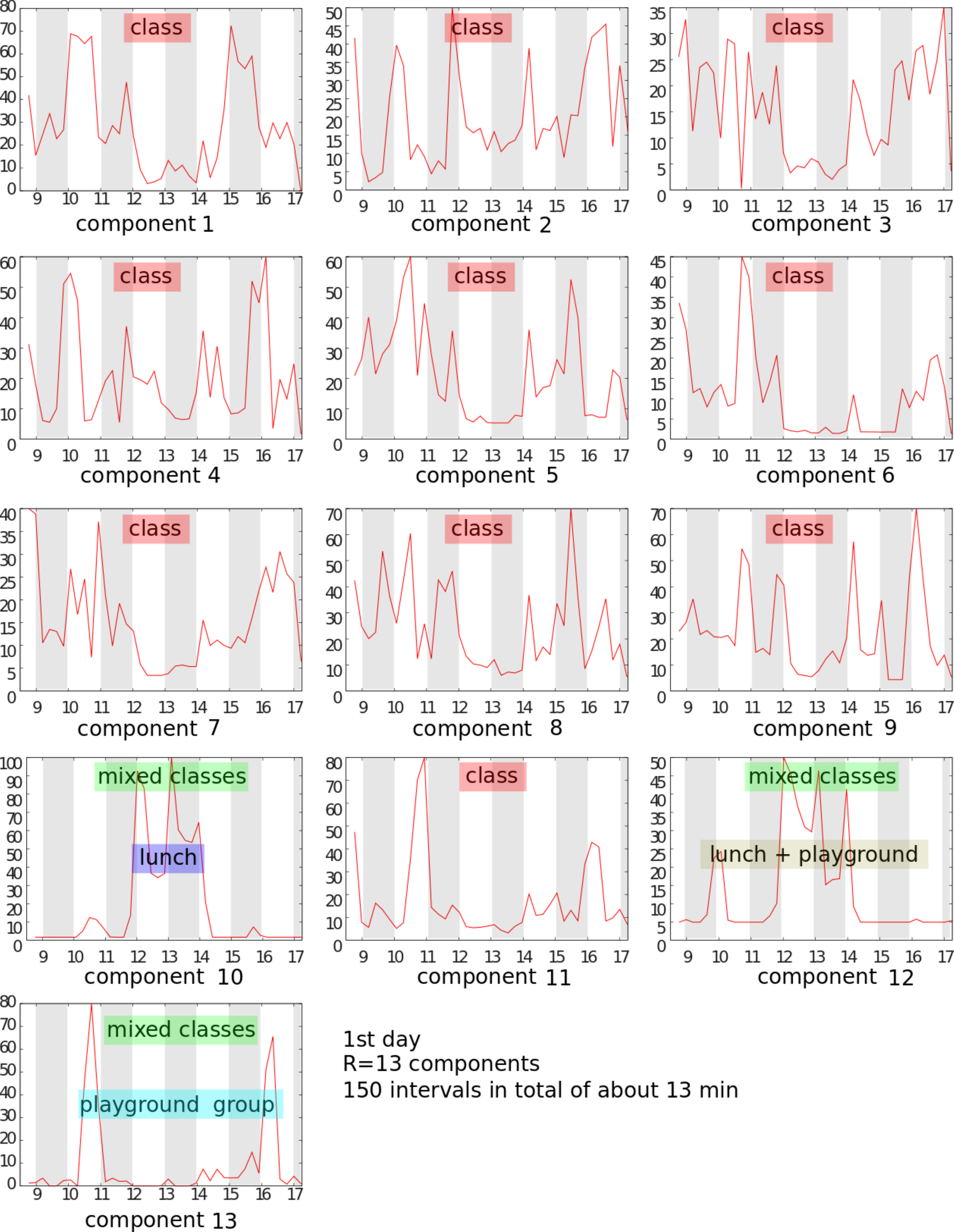}
\end{center}
 \caption{{\bf Activity patterns of the extracted components.}
 Each panel corresponds to one component obtained by non-negative tensor factorization
 of the school temporal network, with $R = 13$, and provides the activity level of the component
 as a function of the time of the day. For clarity, the panels only show the activity patterns for the first day of data (see Fig.~S2 Supplementary Information for the second day).
 Components that can be matched to classes are marked as \textit{class}.
 The other three components that correspond to mixed classes exhibit activity patterns
 that can be understood in terms of gatherings in the social spaces of the school.
 }
\label{Fig:temp}
\end{figure}

As discussed in the Material and Methods section, the factors $\mathbf{C}$ and $\mathbf{A}$ can be combined to compute the activity profiles $s_{kr}$ for each component $r$ as a function of time (i.e., index $k$).
The resulting activity patterns for the $13$ components of our case study are reported in Fig.~\ref{Fig:temp}. For the sake of readability, we show the activity patterns restricted to the first day of the school dataset, only. 
Each panel in figure displays the activity level $s_{kr}$ of an extracted component as a function of the time of the day, from morning to evening. The components are numbered according to the order of Fig.~\ref{Fig:Com} (left to right).
On visual inspection, two main activity patterns can be seen for the extracted components: either the activity is concentrated during class times, with a dip during lunch hours (12pm-2pm), as seen for components $1$-$9$,
or the activity peaks during lunch hours, as seen for component $10$ and $12$.
The mutually disjoint components corresponding to the blocks in Fig.~\ref{Fig:Com}
display the former patterns while the overlapping components $10$ and $12$ display the second pattern.
In all cases, activity levels exhibit large fluctuations over time.
In the following sections we will validate these patterns by mining for the correspondence between the extracted components and the available metadata on the temporal network we study.

\subsection*{Structural validation}
An important peculiarity of the dataset we use is that a ground truth for several important structures is available from node metadata and known activity schedules. Here we validate the community structure extracted by means of non-negative tensor factorization by using the class labels we have for each node, which provide a ground truth on the class structure of the school population.
We want to assess to which extent the components found through factorization correlate with actual classes,
as a function of the number of components $R$.
In order to carry out the validation, we need to match (when possible) the extracted components to the school classes,
and then we proceed to quantify how much of known class structure is recalled, and the corresponding accuracy.
In order to match components to classes, we proceed as follows: as discussed above, for each component $r=1, \dots , R$ of a factorization with $R$ components, we classify the networks nodes as belonging or not belonging to $r$. 
Then we compute the Jaccard overlap between the set of nodes of component $r$ and the set of nodes corresponding to the known $10$ classes (the Jaccard overlap of two sets is the cardinality of their intersection divided by the cardinality of their union), obtaining for each component a vector with the $10$ overlap scores with each known class. When such a vector has only one non-zero value, the corresponding component $r$ is said to match one known class.

Table~\ref{Table_Mod1} reports our results for a number of components ranging from $R=2$ to $R=20$.
For a given number of components $r$ we report the core consistency metric, the number of matched classes/components,
the fraction of nodes spanned by the matches components with respect to the known number of nodes belonging to the matched classes, and other metrics described in the figure caption.
For small values of $R$, the extracted structures communities (classes or mixed classes) of the dataset,
but they only cover part of the network's nodes: only the most prominent set of nodes (in terms of size, presence, connectivity)
are initially uncovered. As $R$ increases, the number of components that can be matched to the classes increases and finally reaches (for $R=12$) the total number of classes of the school.
We remark that the criteria for matching we use (a vector of Jaccard indices with a single non-zero component) is extremely strict,
and yet the factorization technique recovers communities that can be matched to classes for any number of components,
and when a match is achieved, the attribution of nodes to classes is almost perfect, as seen in the table.
In fact, for all choices of $R$, approximately $99\%$ of the nodes that are part of the extracted components
are assigned to the right known community (class). The missing (not assigned) fractions
typically exhibits weak interaction patterns with the rest of the nodes that make their class association
behaviourally ambiguous (we notice that this can arise because of improper sensor behavior or participant compliance). 
Overall, non-negative tensor factorization applied to the adjacency tensor affords an extremely accurate
recovery of the independently known class structure, with a coverage that increases with the number of components $R$
and ultimately recalls almost perfectly all the $10$ known classes.
We remark that for a number of components which is too small to capture the existing class structures,
the technique does not yield partial classes, but rather returns a fewer number of class communities,
or mixed class communities, with high accuracy.

To illustrate the fact that our methodology is efficient at the level of the individual classes, we focus on the case $R=\hat{R}=13$ 
and we report in Table~\ref{table:numbers_com_check} the number of nodes recovered in each of the $10$ mutually disjoint communities that can be matched to the known classes: There is a perfect matching between $9$ components and $9$ classes,
and for the remaining classe there is one student (out of more than $20$) who is not assigned to
the component even though they are known to be part of the class that component represents.
The components that can be matched to classes are marked as ``class'' in Fig.~\ref{Fig:temp}.

\begin{table}[!htbp]
\caption{{\bf Class structure recovered by non-negative tensor factorization as a function of the number of components.}
}
\centering
\begin{tabular}{|l|c||c|c|c|c|c|c|c|}
\hline
$\#$ components & core consistency & $\#$ classes & $\#$ nodes & $\#$ covered & recall \\
\hline
2 & 99.966 & 1 &  25 & 25& 1\\
3 &  99.885 & 1 &  25 & 25& 1 \\
4 & 97.363 & 1  &  25 & 25& 1\\
5 & 96.984 & 2  & 46 & 46 & 1\\
6 & 98.134 & 4  & 94 & 91 & 0.99386\\
7 & 96.604 & 4  & 94 & 91 & 0.99386 \\
8 & 95.451 & 7 & 163 & 162 & 0.99386 \\
9 & 95.408 & 8  & 185 & 184 & 0.99459 \\
10 & 94.815 & 9  & 208 & 207 & 0.99519 \\
11 & 92.639 & 9  & 208 & 207 & 0.99519 \\
12 & 92.526 & 10  & 231 & 230 & 0.99567 \\
13 &  93.472 & 10  & 231 & 230 & 0.99567 \\ 
14 & 78.326 & 10  & 231 & 230 & 0.99567 \\
15 & 76.683 & 10 & 231 & 230 & 0.99567 \\
16 & 73.5 & 10  & 231 & 230 & 0.99567 \\
17 & 65.185 & 10 & 231 & 230 & 0.99567 \\
18 & 60.981 & 10 & 231 & 230 & 0.99567\\
19 & 56.218 & 10 & 231 & 230 & 0.99567\\
20 & 51.709 & 10 &  231 & 230& 0.99567\\
\hline
\end{tabular}
\begin{flushleft}
Each row of the table corresponds to one tensor decomposition, with a number of components ranging from $R=2$
to $R=20$. For each decomposition, we report the core consistency, the number of classes that can be matched to the extracted components, the total number of nodes (students) 
these classes are known to comprise, the number of nodes that belong to the component (i.e., the class nodes covered by the component), and the ratio between the latter two columns, that is, 
the fraction of the known class structure that the method has recalled.
\end{flushleft}
\label{Table_Mod1}
\end{table}

\begin{table}[!htbp]
\caption{{\bf Components vs school classes for $R=13$.}}
\begin{center}
\begin{tabular}{ | l || c | c | c | c | c | c | c | c | c | c| }
\hline
component & 1 & 2 & 3 & 4 & 5 & 6 & 7 & 8 & 9 &11\\
\hline 
component size & 26 &  22 &   23    & 25   &   27   &  23   &   24   &  22  &  24 &  24\\
\hline
class & 1B &    5A  &   3B   &  5B   &   2B &    2A   &  3A  &  4A  &    1A    &       4B\\
\hline 
class size (without teacher) &25   &  22  &  22    & 24  &   26    & 22    &  23  &  21   &  23   &        23\\
\hline 
teacher found &1 & 1& 1& 1 &1 &1 &1 &1&  1& 1\\
\hline 
missing nodes& 0 &1& 0& 0& 0& 0 &0 &0 &0& 0\\
\hline
\end{tabular}
\begin{flushleft}
 Comparison between the extracted components and the class they were matched to, for the case $R=13$.
Missing nodes are nodes (students) that are known to be in the class but do not belong to the components
that was matched to that class.
\end{flushleft}
\end{center}
\label{table:numbers_com_check}
\end{table}

\subsection*{Spatio-temporal validation}
We notice that for $R=13$, three of the extracted components ($10$, $12$, $13$)
are not matched to classes, have temporal activity patterns (Fig.~\ref{Fig:temp})
that set them apart from the other class-related components,
and have significant overlap with the known class-related components (Fig.~\ref{Fig:Com}, right panel).
Here we show that these components correspond to social activities
that involve multiple classes and occur at given points in time and in known spaces of the school.
We validate these activity and mixing patterns by using  the independently known spatio-temporal trajectories
of students, inferred from the radio receivers as described in the Material and Methods section.

We remind that the spatio-temporal information is available for each sensor $s$
in the form of  time-varying location fingerprints $(f_{1}^{(s)}(t), ..., f_{n}^{(s)}(t))$, where the index $n$
runs over radio receivers located in known spaces of the school, that cover both classrooms and social spaces
such as the cafeteria and the playground.
We aggregate the location fingerprints over the same $S=150$ time intervals used to define the tensorial representation of the social network data.
We use the spatio-temporal metadata to study the correlation between the temporal activity of a given component
and the spatial location of the nodes it comprises. Specifically, given a set of nodes $C^r$ corresponding to component $r$,
we define an index of co-location for those nodes as the element-wise product of the location fingerprints $f_{i}^{(s)}(t)$
for all the nodes $s \in C^r$, obtaining a co-location vector $T^r_i(t)$. Notice that $T^r_i(t)$ is non-zero if and only if all the nodes of the component $C^r$ are situated in the vicinity of receiver $i$ at time $t$. 

\begin{figure}[!htbp] 
\begin{center}
  \includegraphics[width=\textwidth]{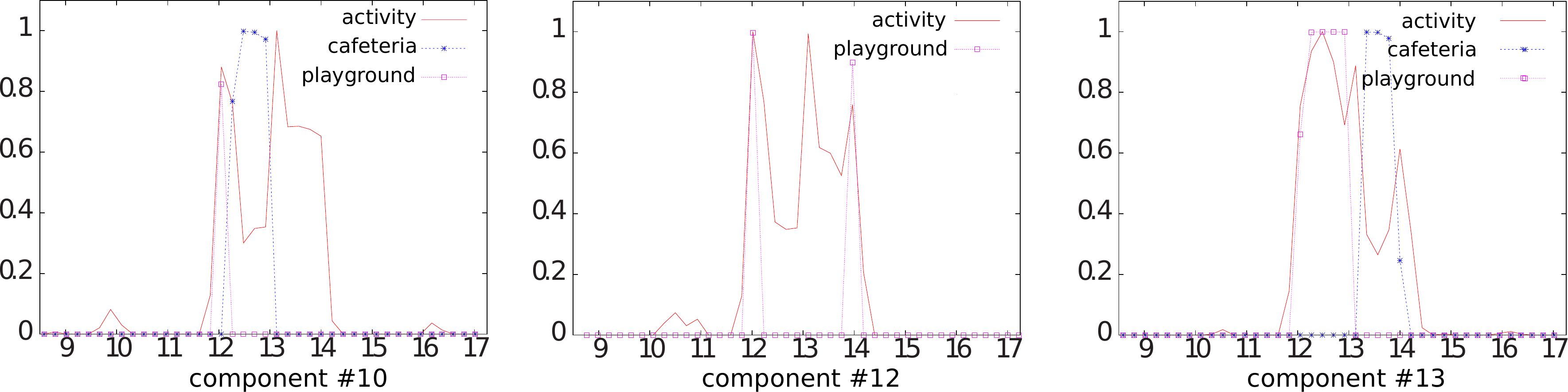}
\end{center}
 \caption{{\bf Activity patterns of components vs co-location in social spaces.}
 Each panel corresponds to one of the three components of Fig.~\ref{Fig:temp}
 that cannot be matched to school classes. The activity pattern of each component
 is compared with the time series of the co-location vector $T^r_i(t)$ ($r=10,12,13$) for two choices of $i$
 that correspond, respectively, to the cafeteria and the playground, i.e., the social spaces of the school.
 The horizontal axis is the time of the day, and the vertical axis has been rescaled for each
 curve so that its maximum is $1$.
 }
\label{mix}
\end{figure}

Finally, we  compare the temporal activity of each non-class component with the co-location vector $T_i(t)$.
On doing so, we observe that activity peaks of each component under study
temporally match the spikes in the corresponding co-location vector for indices $i$
that correspond to locations used for social activities such as the cafeteria or the playground.
Figure~\ref{mix} displays the activity patterns of components $10$, $12$ and $13$
together with the time series of the co-location vector $T^r_i(t)$ for the known social spaces in the school.

The times of these social events, as inferred from non-negative tensor factorization, match the independently known school schedule,
and the classes spanned by the mixed component match the classes that are known to be involved in the social gathering according to the schedule.
We remark that the co-location vector is a very simple and strict way to quantify spatial and temporal co-presence of nodes, and that despite this the results we obtain make the non-class components fully understandable in terms of spatio-temporal coincidences.

In summary, those components that cannot be validated as known classes of the school can be validated in terms of correlated activity patterns, i.e., spatial and temporal coincidences, that are determined by the school activity schedule.

\subsection*{Comparison with community detection algorithms for static networks}
In the above sections we focused on decomposing the tensor representation of a time-varying network
and on separately validating the obtained components in terms of known network communities and temporal activity patterns.
Here we focus on the community structure alone: we compare the communities yielded by non-negative tensor factorization
of the temporal network with those obtained by using well-known community detection algorithms.
Since most community detection algorithms are designed to work on static networks,
we build a static representation of the temporal network by aggregating the time-varying network over time.
The adjacency matrix of the time-aggregated network is defined as
\begin{equation}
t^{(a)}_{ij} = \sum_k \, t_{ijk} \, ,
\end{equation}
i.e., it describes a weighted network where the weight of link $i$-$j$ is the total number of time intervals
during which that link was active (which is proportional to the cumulated duration of the contacts between $i$ and $j$).
We show that our non-negative tensor factorization approach, operating on the time-varying network,
is able to detect the community structure with a performance that is in line with state-of-the-art community detection algorithms
operating on the time-aggregated network.

We regard the known class structure of the network as a ground truth for the community structure,
and evaluate the different methods by testing for the correct assignment of students to classes.
To this end, we define a reference adjacency matrix $\mathbf{R}^{N\times N_C}$,
where $N$ is the number of nodes and $N_C$ is the number of classes.
We set $r_{ij}$ to $1$ if node $i$ belongs to class $j$, and to $0$ otherwise.
For each community detection method $m$, we generate a node-community (node-component) matrix
$\mathbf{P}^{N\times R}$ that encodes the node composition of the $R$ extracted communities:
$p_{ij}$ is set to $1$ if node $i$ is assigned to community $j$, and to $0$ otherwise.
For each community detection method $m$ we define a score matrix $\mathbf{W^m} = \mathbf{R}^T \mathbf{P}$, i.e.,
\begin{equation}
w^{m}_{ij} = \sum_{n=1}^{N} {r_{ni} p_{nj}} \,\, .
 \label{score_matrix}
\end{equation}
This is a $N_C \times R$ matrix, where $w_{ij}$ is the number of nodes in class $i$ that are assigned to community $j$.
The product $\mathbf{R}^T\mathbf{R}$ is the reference score matrix shown in Table~\ref{table:Matrix}:
it is a diagonal $N_C \times N_C$ matrix where the diagonal values are the correct number of students in each class.

For each community detection algorithm $m$ the score matrix $\mathbf{W^m}$ describes the relation between
the obtained communities and the known classes.
If all the detected communities correspond to actual classes,
there is a permutation of the columns of $\mathbf{W^m}$ that makes it diagonal, as in Table~\ref{table:Matrix}.
If a community corresponds exactly to one class, the corresponding value is the matrix diagonal
is the same as in the reference score matrix.

Infomap~\cite{INFOMAP} and the Community Walktrap~\cite{Walktrap} algorithms both yield an exact match,
i.e., the score matrix for these algorithms has a column permutation that is identical to the reference score matrix.
The non-negative tensor factorization approach yields a matrix with a diagonal block (under permutation),
showing that students are correctly attributed to their classes, plus an additional part that describes mixed classes.
In this case one student was not associated to any component,
and consequently one of the communities is smaller than the corresponding class
(see Tables \ref{table:Matrix} and \ref{table:Matrix_NTF}). 
Finally, the Oslom~\cite{OSLOM} and Louvain~\cite{Louvain} algorithms merge several classes into a single community,
and the corresponding score matrices are not diagonal, as shown in the Supplementary Information.

\setcounter{MaxMatrixCols}{20}
\begin{table}
\caption{\textbf{Reference score matrix containing the correct number of students in each class.}}
\begin{center}
$
\begin{pmatrix}
23 & 0 & 0 & 0 & 0 & 0 & 0 & 0 & 0  &0 \\
0& 25 & 0  &0 & 0 & 0&  0 & 0&  0 & 0 \\
 0 & 0 &22 & 0 & 0 & 0&  0 & 0 & 0&  0 \\
 0 & 0 & 0 &26 & 0 & 0 & 0 & 0 & 0&  0 \\
 0 & 0 & 0 & 0 &23 & 0 & 0 & 0 & 0&  0\\
 0 & 0 & 0 & 0 & 0 &22 & 0 & 0 & 0&  0 \\
 0 & 0 & 0 & 0 & 0 & 0 &21&  0 & 0&  0 \\
 0 & 0 & 0 & 0 & 0 & 0 & 0 &23 & 0&  0 \\
 0 & 0 & 0 & 0 & 0 & 0 & 0 & 0& 22&  0 \\
 0 & 0 & 0 & 0 & 0 & 0 & 0 & 0& 0&  24\
 \end{pmatrix}
$
\end{center}
\label{table:Matrix}
\end{table}

\setcounter{MaxMatrixCols}{20}
\begin{table}
\caption{{\bf Matrix $\mathbf{W^\textrm{NTF}}$ obtained through NTF containing the number of students in each community projected over the different classes.}}
\begin{center}
$
  \begin{pmatrix}
23 &  0  & 0  & 0  & 0  & 0  & 0 & 0  & 0  & 0  & 0 &13  & 0\\
  0  &25  & 0  & 0  & 0  & 0  & 0 & 0  & 0  & 0 &  0 &15 & 14\\
  0  & 0  &22  & 0  & 0  & 0  & 0 & 0  & 0  & 0 &  0 &12 &  0\\
  0  & 0  & 0 & 26  & 0 &  0  & 0 & 0  & 0  & 0  & 0 &16  & 0\\
 0   &0  & 0  & 0& 23  & 0  & 0 &  0 &  0  & 0 &23 & 17   &0\\
  0  & 0  & 0  & 0  & 0 & 22 & 0  & 0   &0  & 0 &22  & 9  & 2\\
  0  & 0  & 0  & 0 &  0  & 0  &21  & 0  & 0  & 0  & 0  & 3  &12\\
  0  & 0 &  0  & 0  & 0  & 0  & 0  &23 &  0  & 0 &  0  & 1 &  8\\
  0  & 0  & 0  & 0  & 0  & 0 &  0  & 0 & 21  & 0 &  0   & 1 & 12\\
  0  & 0  & 0  & 0  & 0  & 0  & 0 &  0  & 0  &24   &0  & 1 & 13
  \label{Mat_NTF}
    \end{pmatrix}
$
\end{center}
\label{table:Matrix_NTF}
\end{table}

\section*{Discussion}

We investigated the use of established non-negative tensor factorization techniques
for the detection of the community-activity structure of temporal networks.
The approach we propose is intrinsically temporal and allows
to simultaneously identify network communities together with their activity patterns over time.
Given the lack of widely accepted benchmarks for detecting the community structure of temporal networks,
we evaluated the method by focusing on the special case of an empirical temporal social network
for which we have a ground truth that allows us to validate the structures we detect
in terms of known groups and known activity schedules.
The case we study is a time-varying social network measured in a school by means of wearable sensors:
this is an especially rich and challenging dataset, as it features complex structures
at multiple scales, both topologically and temporally, overlapping communities,
and in general patterns arising from the social and organizational structure of the environment.
We find that non-negative tensor factorization can fully recover the known class structure of the school
and the activity patterns of classes over time. It also yields communities that span multiple classes,
which we validate by using spatio-temporal metadata and link to known social activities in the public spaces of the school.

Detecting temporal network structures by means of non-negative tensor factorization (NTF) provides several advantages.
NTF can naturally deal with the time-varying topology of a temporal network represented as a three-way tensor,
and can yield components that correspond to network communities as well as to correlated activity patterns of network links.
The extracted components/communities can be overlapping, that is, a node can be a member of different components
and the weight of this association is an output of the method and can be used, if necessary, to induce a binary association
between nodes and communities.
The method we study does not depend or rely on temporal continuity: the temporal index of the tensor
is treated as an unordered axis, just like the axes of the adjacency matrices that compose the tensor.
This allows to capture long-range correlations and abrupt changes in the community structure of the network.
The non-negativity constraint affords a simple interpretation of the tensor decomposition
in terms of additive factors that can be linked to known properties or metadata of the system at hand.
The complexity of the model can be tuned to the needs of a specific application or research question
by suitably choosing the number of components used for factorization.
Indicators such as core consistency can be used to assess the robustness of the detected structures
and to diagnose overfitting.
Finally, given the broad use of NTF in surfacing latent signals across a variety of disciplinary domains,
efficient and scalable computational methods for factorization are available.

Several limitations of the described method should be also discussed.
Not relying on the continuity of the tensor along the temporal direction allows to capture global correlations over time
but does not allow to exploit temporal continuity in the (many) cases where continuity is known to be relevant
for the evolution of the network. Properly handling temporal continuity in computing a tensor decomposition
may help to extract robust structures in the presence of noise or missing data.
Exposing and extracting hierarchically-organized or nested community structures
requires to compute multiple tensor factorizations with different numbers of components,
and then to separately establish the correspondence relations or hierarchical relations between
the obtained components.

Possible extensions of the method discussed here include supporting directed temporal networks and weighted temporal networks.
The former is a straightforward extension of the approach presented here. Incremental techniques on non-negative tensor factorization could be developed to deal with real time evolution of the size of the population studied.
Non-negative tensor factorization could be also used to detect structures in multiplex (multi-layer) networks,
which can be equally represented as three-way tensors in which the temporal dimension is replaced by
the index of the network layer.

Finally, we close by highlighting the need for benchmark datasets, containing known synthetic structures,
that could be used to systematically characterize the behavior of different structure detection methods for temporal networks.
The broad availability of empirical temporal network data with a ground truth is also a key enabling
factor for advancing the state of the art in detecting community structures and activity patterns in temporal networks.

\section*{Acknowledgments}
The Authors acknowledge partial support from the Lagrange Project funded by the CRT Foundation, the Q-ARACNE project funded by the Fondazione Compagnia di San Paolo, and the FET Multiplex Project (EU-FET-317532) funded by the European Commission. The Authors thank the French partners of the SocioPatterns collaboration for privileged access to the data used in this study. The Authors acknowledge stimulating discussions with Andrea Martini.

\section*{Author Contributions}
Conceived and designed the experiments: LG AP CC. Performed the experiments: LG. Analyzed the data: LG AP CC. Wrote the paper: LG AP CC.


\bibliographystyle{PLoS}

\newpage
 \renewcommand{\thefigure}{S\arabic{figure}}
 \renewcommand{\thetable}{S\arabic{table}}
\setcounter{figure}{0}
\setcounter{table}{0}
\section*{Supplementary information}

\subsection*{Dataset details}
Details on the measurement of the temporal social network we use are available at the following URL:\\
\url{http://www.sociopatterns.org/publications/}\\
\url{high-resolution-measurements-of-face-to-face-contact-patterns-in-a-primary-school/}
\\
\\
The aggregated (static) version of the network, together with node metadata are available at:\\
\url{http://www.sociopatterns.org/datasets/primary-school-cumulative-networks/}
The table \ref{table:class} yields the number of students in each class of the school.
\begin{table}[!htbp]
\caption{{\bf Number of students in each class of the school.}}
\begin{center}
\begin{tabular}{ | c | c |  }
\hline 
class & class size\\
\hline 
1A &  23 \\
1B & 25 \\
2A  & 22 \\
2B & 26 \\
3A  & 23 \\
3B  & 22 \\
4A  &  21\\
4B & 23 \\
5A& 22\\
5B & 24 \\
\hline
\end{tabular}
\end{center}
\label{table:class}
\end{table}

\subsection*{Components and activity patterns}
\begin{figure}[!ht]
\begin{center}
\includegraphics[width=0.85\textwidth]{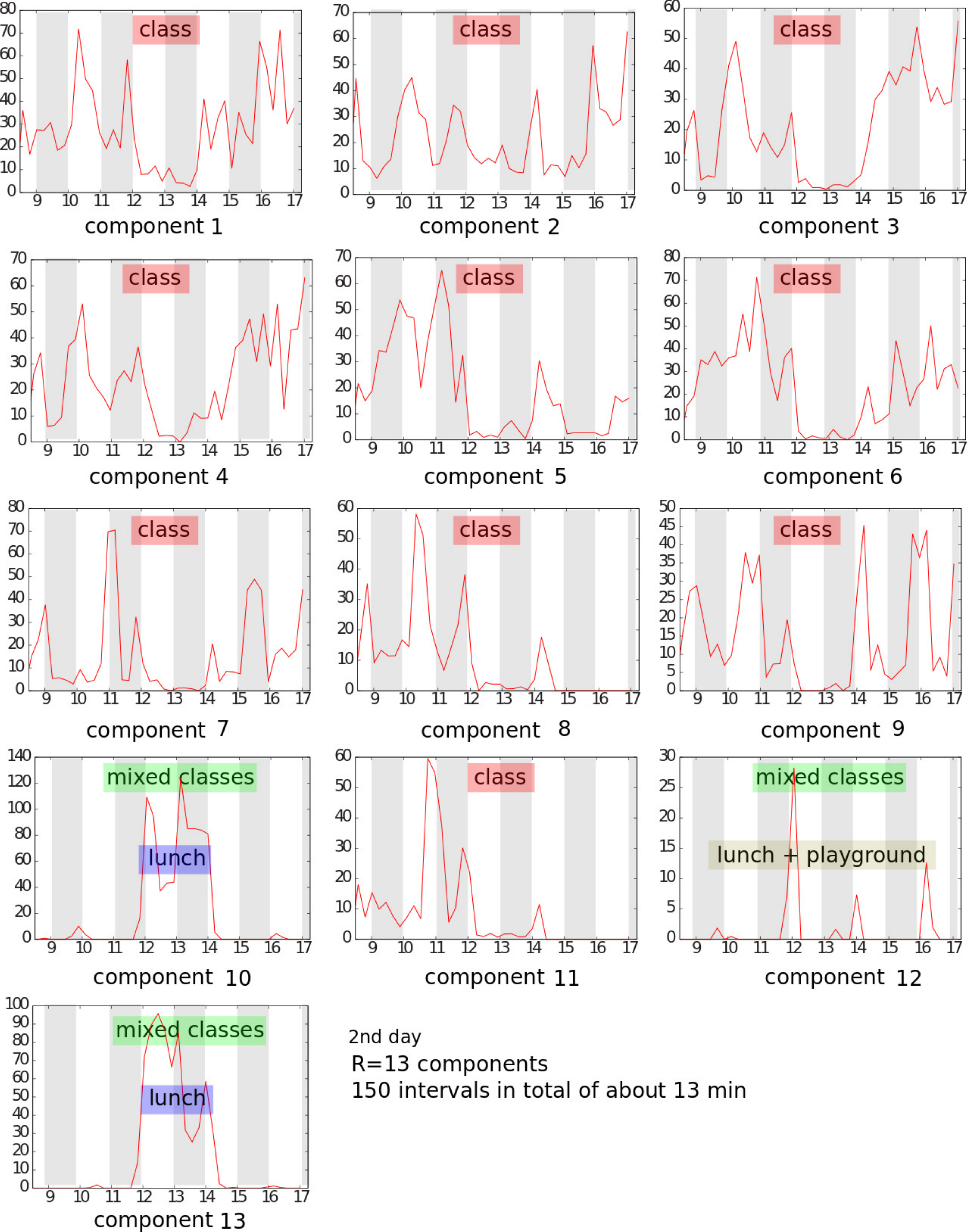}
\end{center}
\caption{
{\bf Activity patterns of the extracted components, second day.}
 Each panel corresponds to one component obtained by non-negative tensor factorization
 of the school temporal network, with $R = 13$, and provides the activity level of the component
 as a function of the time of the day. Components that can be matched to classes are marked as \textit{class}.
 The other three components that correspond to mixed classes exhibit activity patterns
 that can be understood in terms of gatherings in the social spaces of the school.
}
\label{fig:panel_day2}
\end{figure}

The non-negative tensor factorization with $13$ components has been applied on a tensor built from a temporal network of contacts in a primary school. The generic representation of the
factorization is displayed on Fig.\ref{fact_result}. The $13$ components have different sizes summarized in Tab.\ref{table:numbers_com}. The activity patterns of the components during the second day of the experiment
are displayed in Fig.\ref{fig:panel_day2}.
\begin{figure}[!ht] 
\begin{center}
     \includegraphics[width=7cm]{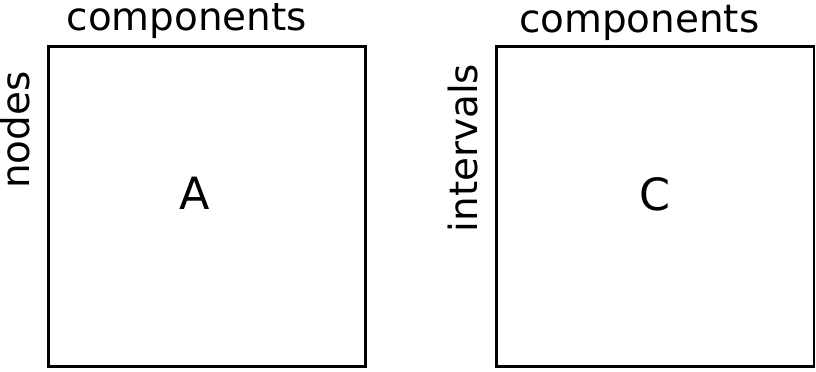}
\end{center}
 \caption{{\bf Schematic representation of the factorization result for an undirected temporal network.}
 The factors $\mathbf{A}$ and $\mathbf{C}$ are matrices with $R$ columns, each correspond to one extracted component.
 The rows of $\mathbf{A}$ correspond to network nodes, and the rows of $\mathbf{C}$  to discrete time intervals.
 The entries of $\mathbf{A}$ give the membership weight of nodes to the different components.
 The entries of $\mathbf{C}$ give the activity level of components at different points in time.
 }
\label{fact_result}
\end{figure}
\begin{table}[!ht]
\caption{{\bf Size of the components extracted by non-negative tensor factorization with $R=13$.}}
\begin{center}
\begin{tabular}{ | c | c |  }
\hline 
community & community size\\
\hline 
1 &  26 \\
2 & 22 \\
3  & 23 \\
4 & 25 \\
5  & 27 \\
6  & 23 \\
7  &  24\\
8 & 22 \\
9& 24\\
10 & 89 \\
11 & 24 \\
12& 61\\
13 & 46 \\
\hline
\end{tabular}
\end{center}
\begin{flushleft}
The size is computed after carrying out a binary classification of the nodes
as belonging or not belonging to a given component, on the basis of the membership weights.
\end{flushleft}
\label{table:numbers_com}
\end{table}

\subsection*{Tuning the complexity of the model}
\begin{figure}[!ht]
\begin{center}
\includegraphics[width=0.9\textwidth]{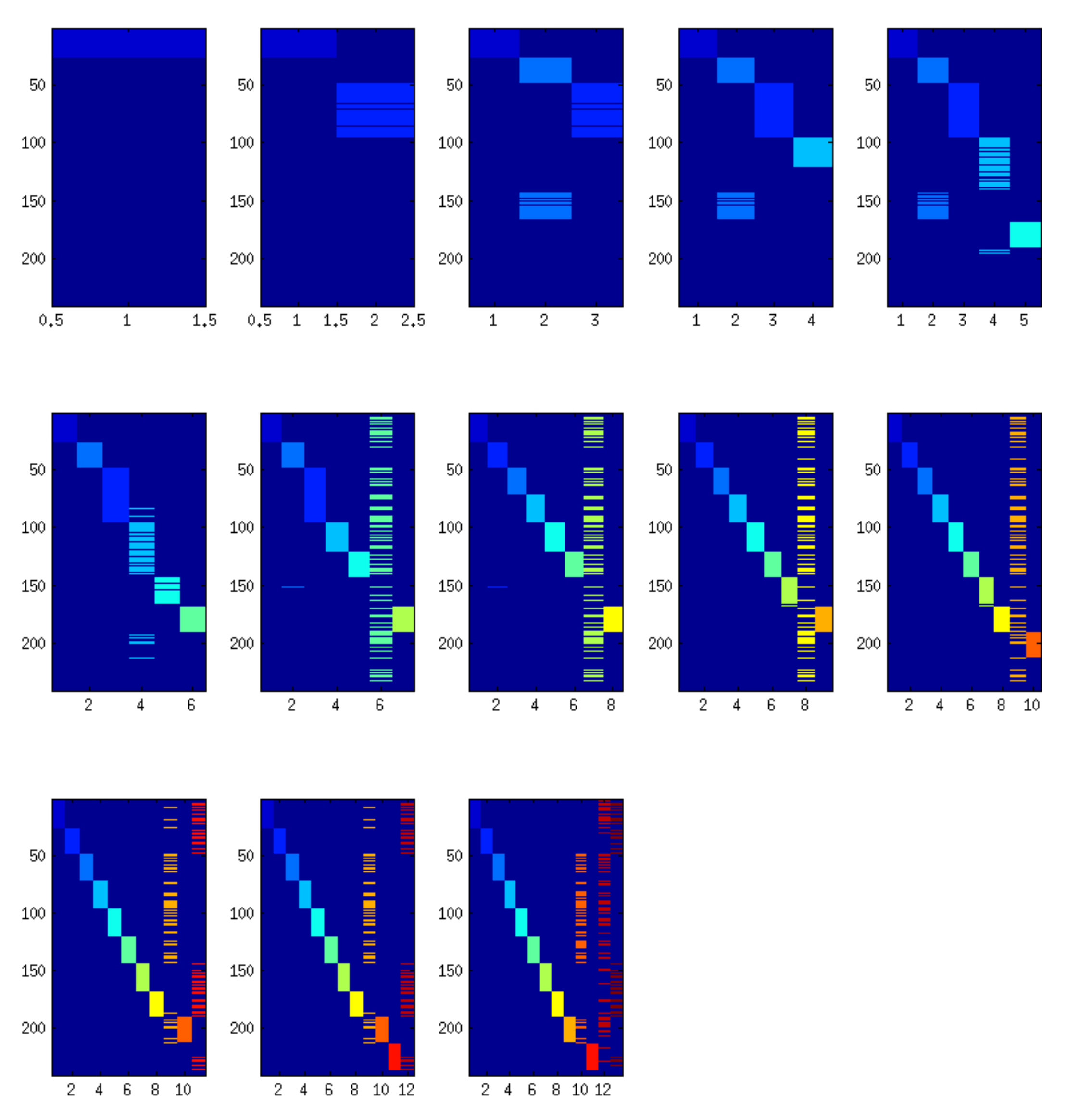}
\end{center}
\caption{{\bf Component-node matrix for the number of components ranging from $R=1$ to $R=13$.}
Rows correspond to network nodes and columns to components. The matrix is obtained from the factor $\mathbf{A}$ by classifying each node as belonging (lighter rectangles) or not belonging (dark blue rectangles) to a given component. Colors identify components. The order of the nodes is the same in all the subplots.
}
\label{fig:Rscan}
\end{figure}

Here, we assess how the detected community structure evolves with the number of components $R$, using the case $R=13$ (discussed in the main text) as a reference. To assess the consistency of the detected structure across different decompositions,
we display in Fig.~\ref{fig:Rscan} the  component-node matrix for different values of $R$ ,  keeping the order of the nodes unchanged across all plots. We observe that for small values of $R$ the detected communities are in general larger than for $R=13$: this is due to the fact that factorization attempts to describe as much of the temporal network as possible, and returning small components would be sub-optimal.
For small values of $R$, when two or more classes are found to be merged in one component, the merged classes are consistently those that participate in the overlapping communities found for $R=13$. An example of this can be seen in the case $R=5$, where two components correspond to classes, while the three other components each mix two classes that participate in the same overlapping communities seen for $R=13$.

\subsection*{Delineation of communities}
\begin{figure}[!ht]
\begin{center}
\includegraphics[width=0.7\textwidth]{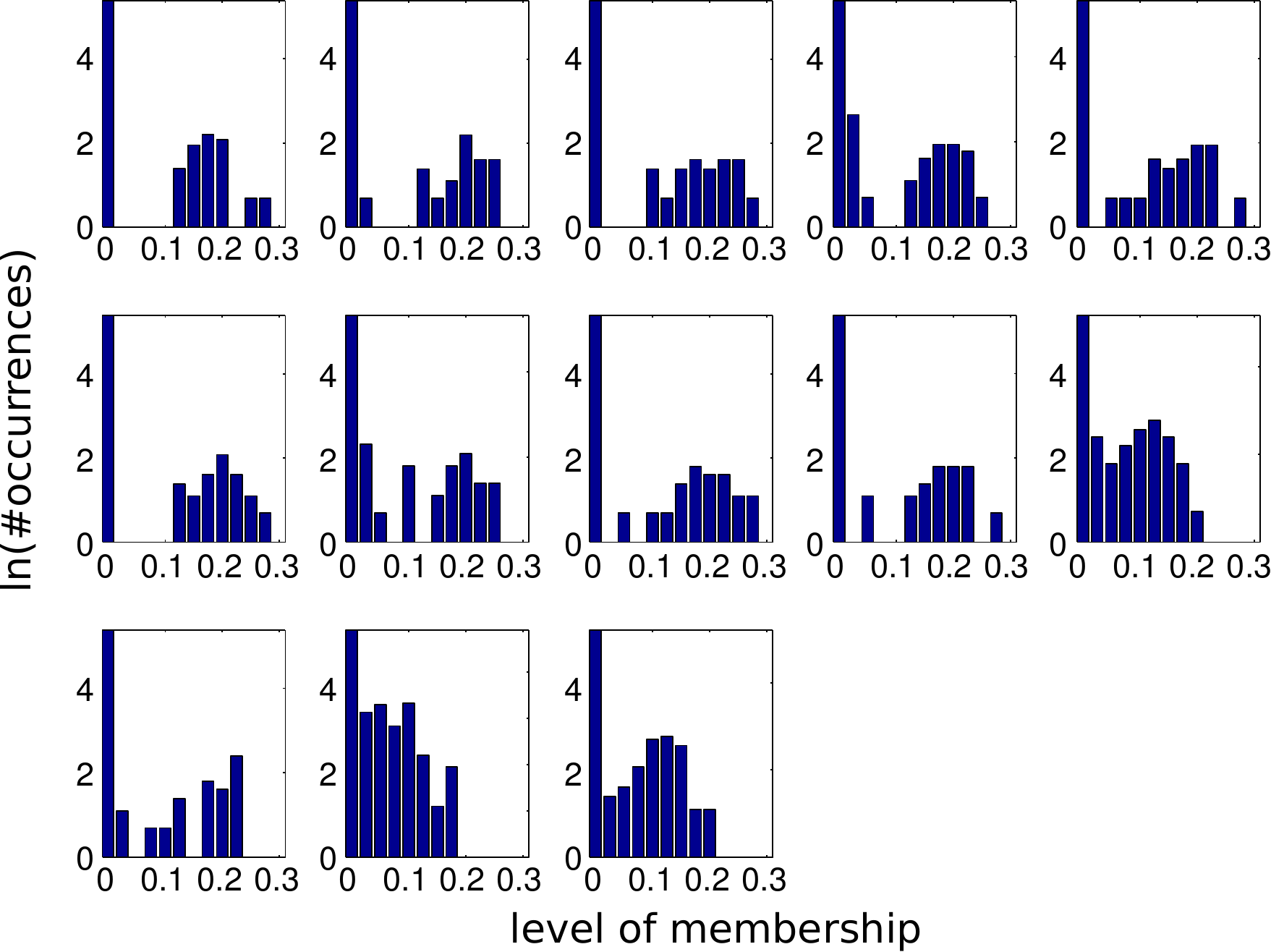}
\end{center}
\caption{
{\bf Histograms of membership weights for $R=13$ components.}  
}
\label{Fig:memb_ln}
\end{figure}

\begin{figure}[!ht]
\begin{center}
\includegraphics[width=0.6\textwidth]{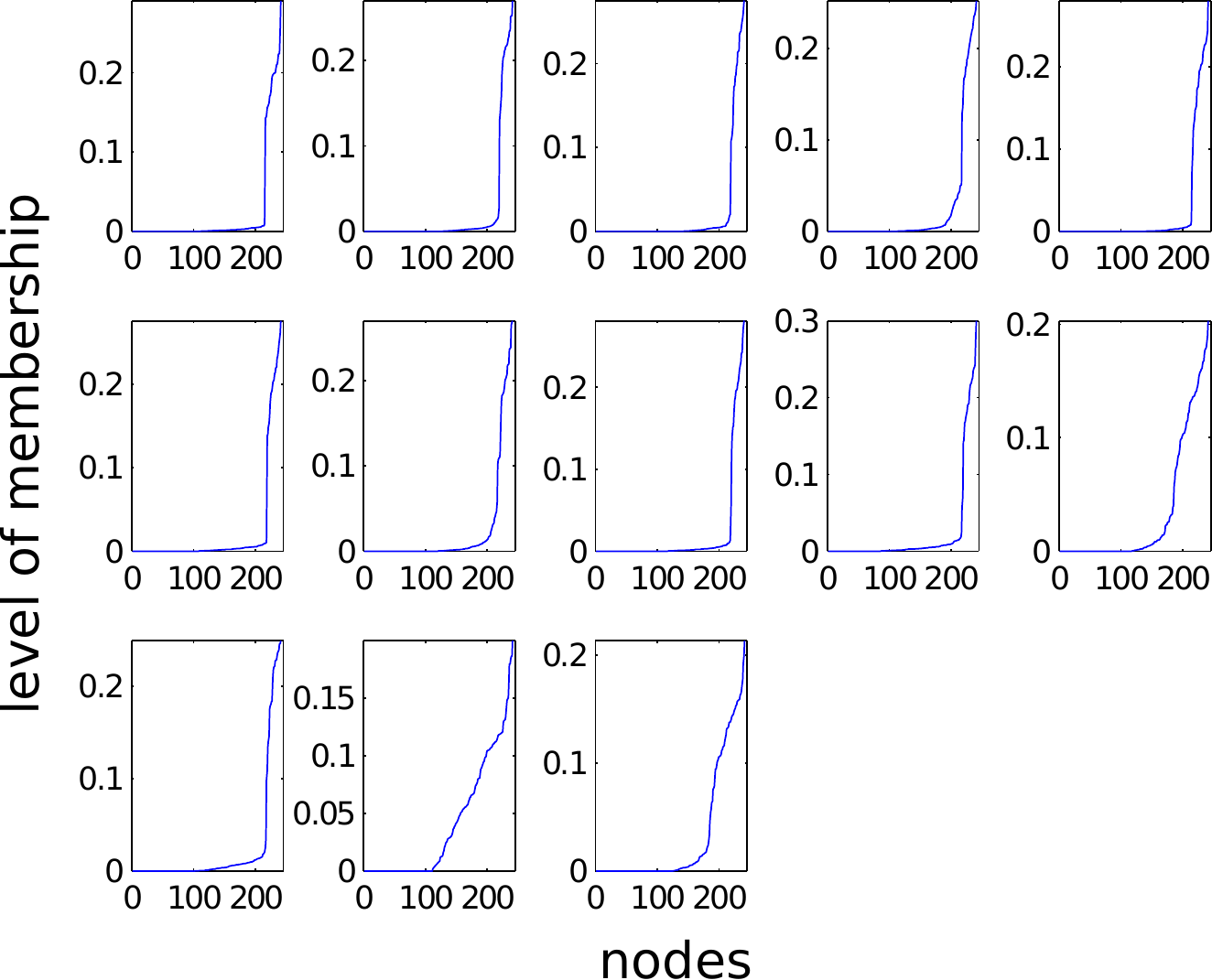}
\end{center}
\caption{
{\bf Weight-rank plots of membership weights for $R=13$ components.}  
}
\label{Fig:ranked_weights}
\end{figure}

For the case $R=13$ discussed in the main text, Fig.~\ref{Fig:memb_ln} reports the distribution of the membership weights for each component. For all components, a large fraction of nodes have zero weights (notice that we plot the natural logarithm of the weights).
This is also visible in Fig.~\ref{Fig:ranked_weights}, where the weights for each components are ranked and plotted as a function of rank.

\subsection*{Robustness of the detected structures with respect to snapshot duration}
All the results presented in the main text were obtained using a $13$ minutes aggregation interval. To choose the appropriate aggregation level, one has to consider the characteristic time scale of the events occurring in the network (if there is any).
If the chosen time scale is too short, tensor factorization may return components that represent noise,
whereas the user of a time scale which is too long may blur the temporal features of the temporal network.
The $13$ minutes time interval we used in the main text is a natural time scale to resolve the scheduled school activities,
but it is nevertheless a somehow arbitrary choice. Therefore we assess the stability of our results with respect
to the duration of the temporal aggregation interval by repeating our analysis for interval durations of $5, 15, 30, 60$ minutes.
In all cases, we fixed $R = 13$ components.

\begin{figure}[!htbp]
\begin{center}
\includegraphics[width=0.7\textwidth]{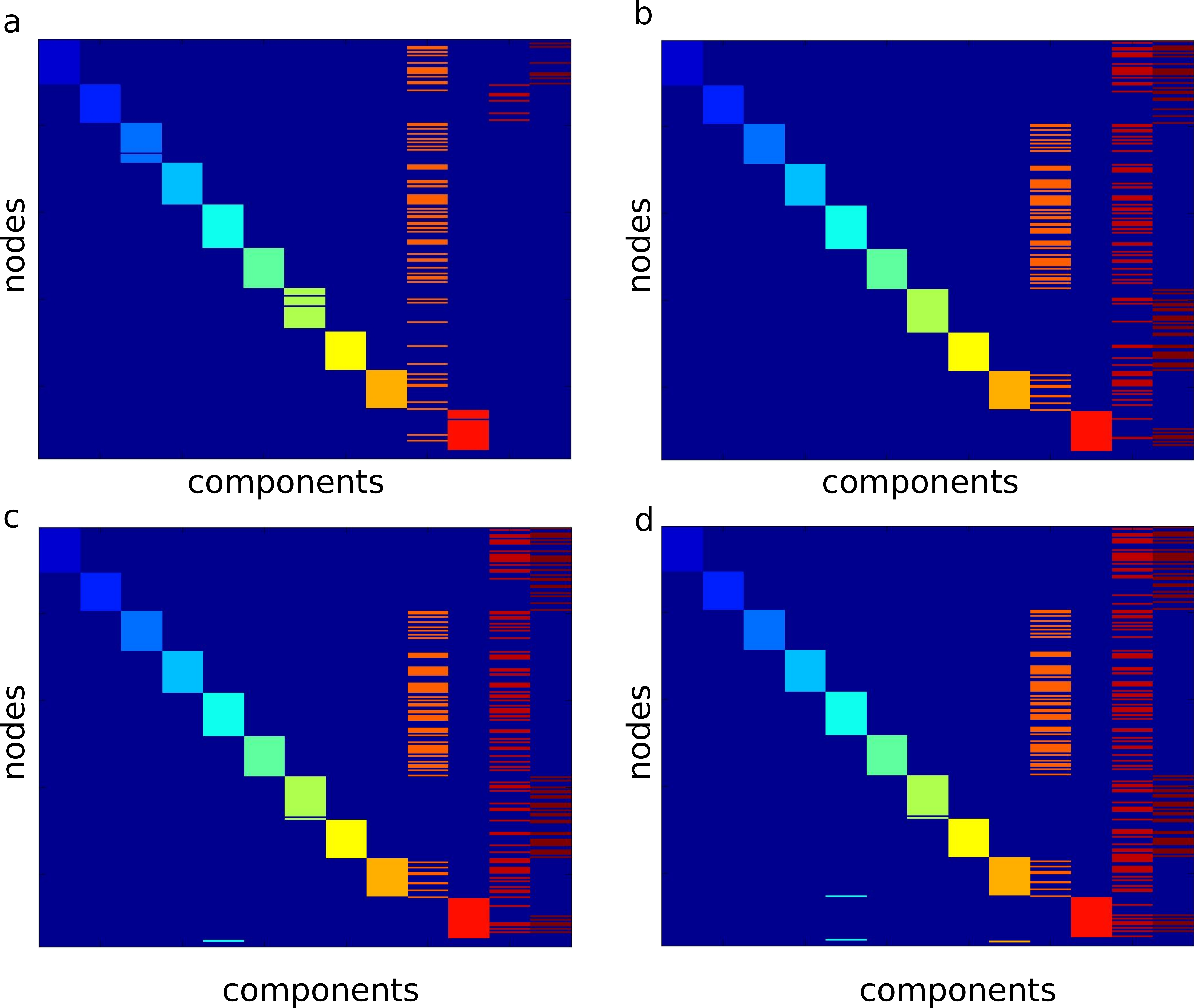}
\end{center}
\caption{\bf{Component-node matrix for $R=13$ components, for different granularities}. a) $5$ min, b) $15$ min, c) $30$ min and d) $60$ min. Rows correspond to network nodes and columns to components.}
The matrix is obtained from the factor $\mathbf{A}$ by classifying each node as belonging (lighter rectangles) or not belonging (dark blue rectangles) to a given component. The order of the nodes has been rearranged to expose 
the block structure of the matrix. Colors identify components, and the community structures that can be matched to school classes are annotated with the corresponding class name.
\label{Fig:snapshots_comp}
\end{figure}

Figure~\ref{Fig:snapshots_comp} shows that the general structure of the factor matrices we obtain is very similar:
for all values of the interval duration all the school classes are found. 
To further investigate this point we can look at the number of students assigned to each class,
computing the score matrix for each case.
We show in Tables~\ref{table:Matrix_60}, \ref{table:Matrix_30}, \ref{table:Matrix_15} and \ref{table:Matrix_5}
that the score values are very close to one another, i.e., for this range of time intervals
the structures we detect appear to be robust with respect to changes in the duration of the aggregation interval.

\setcounter{MaxMatrixCols}{20}

\begin{table}[!htbp]
 \caption{{\bf Matrix containing the projection of the $13$ components - obtained through NTF applied on the tensor built with an aggregation of $60$ minutes - over the different classes. }}
\begin{center}
$
  \begin{pmatrix}
 22  &  0   &  0   &  0  &   0  &   0  &   0  &   0   &  0 &    0  &   0  &  12  &   6\\
 0  &   25  &   0  &   0   &  0  &   0   &  0   &  0  &   0   &  0   & 14   & 15  &   0\\
 0    & 0  &  22   &  0    & 0    & 0  &   0   &  0 &    0  &   0   &  0   &  8  &  12\\
 0   &  0   &  0   & 26  &   0   &  0   &  0   &  0  &   0   &  0    & 0    & 11    &14\\
  0    & 0    & 0   &  0    &23    & 0    & 0  &   0    & 0    & 0   &  0  &  11   & 15\\
  0   &  0   &  0   &  0   &  0  &  22   &  0   &  0   &  0    & 0   &  0   &  8   &  8\\
 0   &  0 &    0    & 0   &  0  &   0  &  21  &   0   &  0    & 0   & 12   & 10   &  0\\
 0   &  0   &  0   &  0   &  0  &   0   &  0 &   22   &  0  &   0   &  6  &   8   &  0\\
 0   &  0  &   0    & 0  &   0   &  0  &   0   &  0  &  21    & 0  &  10 &    4  &   0\\
 0   &  0    & 0   &  0   &  0 &    0  &   0   &  0    & 0 &   23 &   13  &   9  &   0
 \end{pmatrix}
  $
\end{center}
\label{table:Matrix_60}
\end{table}

\setcounter{MaxMatrixCols}{20}
\begin{table}[!htbp]
\caption{{\bf Matrix containing the projection of the $13$ components - obtained through NTF applied on the tensor built with an aggregation of $30$ minutes - over the different classes. }}
\begin{center}
$
  \begin{pmatrix}
      22   &  0    & 0  &   0 &    0 &    0 &    0 &    0 &    0 &    0&     0 &   12&     6\\
     0  &  25   &  0   &  0    & 0    & 0   &  0  &   0   &  0    & 0   & 13   & 15  &   0\\
     0  &   0   & 22   &  0    & 0   &  0  &   0   &  0   &  0    & 0   &  0   &  8 &   12\\
     0  &   0   &  0   & 25    & 0   &  0  &   0  &   0   &  0   &  0  &   0   & 12  &  14\\
     0  &   0   &  0   &  0   & 23   &  0  &   0   &  0   &  0   &  0  &   0  &  11 &   14\\
     0  &   0  &   0   &  0  &   0   & 22  &   0   &  0    & 0   &  0  &   0  &   8 &    8\\
     0   &  0  &   0   &  0  &   0  &   0  &  21  &   0   &  0   &  0  &  12  &   5 &    0\\
     0  &   0  &   0  &   0  &   0  &   0  &   0  &  22   &  0   &  0  &   6  &   5  &   0\\
     0  &   0  &   0   &  0  &   0  &   0  &   0  &   0   & 21   &  0  &   9    & 1  &   0\\
     0  &   0   &  0   &  0  &   0  &   0  &   0   &  0   &  0  &  23 &   13  &   7  &   0
  \end{pmatrix}
$
\end{center}
\label{table:Matrix_30}
\end{table}

\setcounter{MaxMatrixCols}{20}
\begin{table}[!htbp]
\caption{{\bf Matrix containing the projection of the $13$ components - obtained through NTF applied on the tensor built with an aggregation of $15$ minutes - over the different classes. }}
\begin{center}
$
  \begin{pmatrix}
     22   &  0  &   0   &  0  &   0  &   0    & 0   &  0  &   0  &   0   &  0 &   11    & 6\\
     0   & 25   &  0  &   0 &    0 &    0 &    0  &   0  &   0  &   0 &   13  &  15 &    0\\
     0   &  0  &  22  &   0 &    0 &    0  &   0     &0   &  0  &   0   &  0    & 8  &  12\\
     0    & 0  &   0   & 24 &    0   &  0  &   0   &  0   &  0   &  0  &   0 &   11   & 14\\
     0    & 0  &   0  &   0  &  23   &  0   &  0   &  0    & 0   &  0  &   0  &  10 &   15\\
     0    & 0  &   0  &   0    & 0   & 22    & 0   &  0   &  0 &    0  &   0  &   8    & 8\\
     0    & 0   &  0     &0  &   0    & 0   & 21   &  0 &    0   &  0 &   12  &   4  &   0\\
     0    & 0   &  0  &   0 &    0  &   0  &   0 &   22   &  0 &    0  &   6  &   2   &  0\\
     0   &  0  &   0   &  0   &  0  &   0   &  0   &  0  &  21 &    0 &    8   &  1 &    0\\
     0   &  0   &  0  &   0   &  0   &  0  &   0   &  0   &  0  &  24  &  13    & 4  &   0
  \end{pmatrix}
$
\end{center}
\label{table:Matrix_15}
\end{table}

\setcounter{MaxMatrixCols}{20}
\begin{table}[!htbp]
\caption{{\bf Matrix containing the projection of the $13$ components - obtained through NTF applied on the tensor built with an aggregation of $5$ minutes - over the different classes. }}
\begin{center}
$
  \begin{pmatrix}
       22  &   0 &    0 &    0 &    0 &    0 &    0 &    0 &    0 &    0&     0 &    0 &    5\\
     0   & 25   &  0   &  0  &   0  &   0  &   0  &   0   &  0   &  0  &   6    & 0  &  12\\
     0  &   0  &  22   &  0 &    0  &   0 &    0  &   0  &   0  &   0  &   0  &   0 &    8\\
     0  &   0 &    0   & 24 &    0 &   0  &   0   &  0    & 0  &   0  &   0   &  0   & 12\\
     0  &   0 &    0  &   0 &   23   &  0  &   0  &   0  &   0 &    0 &    0 &    0  &  12\\
     0 &    0 &    0  &   0 &    0  &  22   &  0  &   0  &   0 &    0  &   0 &    0  &   8\\
     0  &   0&     0  &   0 &    0  &   0  &  21  &   0  &   0  &   0  &   0 &    0  &   2\\
     0  &   0 &    0  &   0 &    0  &   0 &    0  &  21  &   0  &   0  &   0 &    0  &   2\\
     0    & 0  &   0  &   0 &    0  &   0 &    0  &   0  &  21   &  0  &   0 &    5  &   1\\
     0   &  0   &  0  &   0 &    0  &   0 &    0  &   0  &   0  &  21   &  0 &    0 &    3
  \end{pmatrix}
$
\end{center}
\label{table:Matrix_5}
\end{table}

\subsection*{Comparison with reference community detection algorithms}
Here we report the score matrices $W^m$, as defined in the main text, for some well-known community detection algorithms
for static networks. As discussed in the main text, Infomap~\cite{INFOMAP} and the Community Walktrap~\cite{Walktrap} algorithms yield score matrices that are equivalent to the reference score matrix (up to column permutations).
Oslom~\cite{OSLOM} and the Louvain~\cite{Louvain} algorithms, conversely, merge several known classes
into larger communities, thus the corresponding score matrices are not diagonal
(see Tables \ref{table:Matrix_Oslom} and \ref{table:Matrix_Louvain}).

\setcounter{MaxMatrixCols}{20}
\begin{table}
\caption{{\bf Matrix containing the projection of the components - obtained through the OSLOM algorithm applied on the aggregated network - over the different classes.}}
\begin{center}
$
  \begin{pmatrix}
23 & 0 & 0&  0 & 0 & 0 & 0 & 0\\
 0 &25&  0 & 0 & 0&  0 & 0 & 0\\
 0 & 0 &22 & 0 & 0 & 0 & 0 & 0\\
 0 & 0 & 0& 26 & 0&  0 & 0 & 0\\
0 & 0 & 0  &0& 23 & 0&  0 & 0\\
 0 & 0&  0&  0 &22 & 0 & 0 & 0\\
 0 & 0 & 0&  0 & 0 &21 & 0 & 0\\
 0 & 0 & 0&  0 & 0 & 0 &23 & 0\\
 0 & 0 & 0&  0 & 0&  0 & 0 &22\\
0 & 0 & 0&  0 & 0 & 0 & 0 &24
  \end{pmatrix}
$\\
\end{center}
\label{table:Matrix_Oslom}
\end{table}

\setcounter{MaxMatrixCols}{20}
\begin{table}
\caption{{\bf Matrix containing the projection of the components - obtained through the Louvain algorithm applied on the aggregated network - over the different classes.}}
\begin{center}
$
  \begin{pmatrix}
23 & 0 & 0 & 0&  0 & 0\\
  0 &25 & 0 & 0 & 0&  0\\
  0 & 0& 22&  0 & 0 & 0\\
  0 & 0& 26 & 0 & 0 & 0\\
  0 & 0 & 0& 23 & 0 & 0\\
 0  &0 & 0 &22 & 0 & 0\\
  0 & 0 & 0 & 0 &21&  0\\
  0 & 0 & 0 & 0 &21&  2\\
  0 & 0 & 0 & 0 & 0 &22\\
  0 & 0&  0&  0 & 0 &24
   \end{pmatrix}
$\\
\end{center}
\label{table:Matrix_Louvain}
\end{table}

\end{document}